\documentclass[superscriptaddress,amsmath,amssymb,aps,pre,twocolumn,footinbib]{revtex4}

\usepackage[normal]{subfigure}
\usepackage{amsmath}
\usepackage{dcolumn}
\usepackage{bm}

\usepackage{pxfonts}

\usepackage[normalem]{ulem}
\usepackage{graphicx,color}
\usepackage{subfigure}
\usepackage{bbold}
\usepackage{cancel}
\usepackage[colorlinks,citecolor=red,linkcolor=blue]{hyperref}
\usepackage[usenames,dvipsnames,svgnames]{xcolor}
\usepackage{calligra}
\usepackage{mathrsfs}
\DeclareMathAlphabet{\mathcalligra}{T1}{calligra}{m}{n}

\newcommand{\be}{\begin{equation}}
\newcommand{\ee}{\end{equation}}

\newmuskip\pFqmuskip

\newcommand*\pFq[6][8]{%
  \begingroup 
  \pFqmuskip=#1mu\relax
  \mathcode`\,=\string"8000
  \begingroup\lccode`\~=`\,
  \lowercase{\endgroup\let~}\pFqcomma
  {}_{#2}F_{#3}{\left[\genfrac..{0pt}{}{#4}{#5};#6\right]}%
  \endgroup
}
\newcommand{\pFqcomma}{\mskip\pFqmuskip}

\DeclareRobustCommand{\mathbfup}[1]{\begingroup\changegreekbf\mathbf{#1}\endgroup}
\makeatletter
\def\changegreek{\@for\next:={%
  alpha,beta,gamma,delta,epsilon,zeta,eta,theta,kappa,lambda,mu,nu,xi,pi,rho,sigma,%
  tau,upsilon,phi,chi,psi,omega,varepsilon,vartheta,varpi,varrho,varsigma,varphi}%
  \do{\expandafter\let\csname\next\expandafter\endcsname\csname\next up\endcsname}}
\def\changegreekbf{\@for\next:={%
  alpha,beta,gamma,delta,epsilon,zeta,eta,theta,kappa,lambda,mu,nu,xi,pi,rho,sigma,%
  tau,upsilon,phi,chi,psi,omega,varepsilon,vartheta,varpi,varrho,varsigma,varphi}%
  \do{\expandafter\def\csname\next\expandafter\endcsname\expandafter{%
    \expandafter\bm\expandafter{\csname\next up\endcsname}}}}
\makeatother

\begin{document}
\title{Surface tension and a self-consistent theory of soft composite solids with elastic inclusions}

\author{Francesco Mancarella}
\affiliation{Nordic Institute for Theoretical Physics, Royal Institute of Technology and Stockholm University, SE-106 91 Stockholm, Sweden}
\author{John S. Wettlaufer}
\affiliation{Yale University, New Haven, Connecticut 06520, USA}
\affiliation{Nordic Institute for Theoretical Physics, Royal Institute of Technology and Stockholm University, SE-106 91 Stockholm, Sweden}
\affiliation{Mathematical Institute, University of Oxford, Oxford OX1 3LB, UK}
\begin{abstract}
The importance of surface tension effects is being recognized in the context of soft composite solids, where it is found to significantly affect the mechanical properties, such as the elastic response to an external stress. It has recently been discovered that Eshelby's inclusion theory breaks down when the inclusion size approaches the elastocapillary length $L\equiv\gamma/E$, where $\gamma$ is the inclusion/host surface tension and $E$ is the host Young's modulus.   Extending our recent results for liquid inclusions, here we model the elastic behavior of  a non-dilute distribution of isotropic elastic spherical inclusions in a soft isotropic elastic matrix, subject to a prescribed infinitesimal far-field loading. Within our framework, the composite stiffness is uniquely determined by the elastocapillary length $L$, the spherical inclusion radius $R$, and the stiffness contrast parameter ${C}$, which is the ratio of the inclusion to the matrix stiffness.  We compare the results with those from the case of liquid inclusions, and we derive an analytical expression for elastic cloaking of the composite by the inclusions. Remarkably, we find that the composite stiffness is influenced significantly by surface tension even for inclusions two orders of magnitude more stiff than the host matrix. Finally, we show how to simultaneously determine the surface tension and the inclusion stiffness using two independent constraints provided by global and local measurements.
\end{abstract}
\date{\today}
\maketitle
\section{Introduction}

Although the effects of surface tension are commonly studied for fluid/fluid interfaces, recently their importance has been demonstrated in composite soft solids \cite{Style16arxiv, Jerison11, Style12, StylePRL13, StylePNAS13, Chakrabarti13, Mora13, Mora10, Mora11, StyleNatComm13, Style15, MSW3phase, MSWa}.
In general, we associate a surface energy $\gamma$ with any material surface, due to the work required to form it, and a surface stress $\Upsilon$, which opposes the stretching of the surface \cite{deGennes10} and corresponds to the reversible work per unit area required to create new interfacial area by stretching. In general $\Upsilon$ is a symmetric, second order, two-dimensional tensor, which here we treat as a nearly isotropic, strain-independent scalar. Therefore, in this case, $\Upsilon$ coincides with the surface energy $\gamma$.  Hence, both can be referred to as surface tension, and tractions at the two sides of the surface fulfil the generalized Young-Laplace equation $(\sigma_{tot}^{(2)}-\sigma_{tot}^{(1)}) \cdot \textbf{n}=\gamma \mathcal{K} \textbf{n}$ \cite{Style15, Stylesoft15}. Because this boundary condition shows good agreement with the experimental behavior of hydrogels and silicone gels \cite{Mora10,Style15,StylePRL13,StyleNatComm13,Mora11,Jerison11,Mora13,Karpit15}, we regard surface tension as a valid approximation for the interfacial stress of soft materials. The differences between surface stress and surface energy are discussed in detail in Refs. [\onlinecite{Gurtin75, Steinmann08, Javili10, Style16arxiv}].

Depending on the length scales involved, surface tension can have a significant impact on the mechanical properties of both fluids and compliant solids. In liquid systems the principle of minimizing the total interfacial free energy subject to a constraint underlies a large range of capillary phenomena, such as the spherical shape of small droplets and the angle of contact between liquid, vapor and solid phases, as described by the Young-Dupr\'e equation \cite{deGennes10}.  Such basic phenomena inspires microfluidic applications, such as precisely-controlling the actuation of droplets in contact with purposely designed surfaces \cite{Bouasse24, Weislogel97,Bico02,Darhuber05}. Moreover, as noted above, the role played by surface tension in the mechanical behavior of soft composites is an area of rapidly growing interest \cite{Long96, Jerison11, Style12, StylePRL13, StylePNAS13, Chakrabarti13, Mora13}.  Indeed, the influence of surface tension  underlies the formation of ripples and creases on soft solids \cite{Mora10}, and the smoothing of their free surfaces \cite{Mora11, Chen12, Gordan08, Persson10, Jagota12}. Furthermore, accounting for surface tension leads to a revision  \cite{StyleNatComm13} of the classical JKR theory of elastic contact \cite{Johnson71}.  Importantly for our treatment here, accounting for surface tension in composite soft solids with fluid inclusions can lead to {\em either} stiffening {\em or} softening of the composite, in both the dilute \cite{Style15,Stylesoft15} and non-dilute \cite{MSW3phase,MSWa} cases, in agreement with experiments. 

We understand the stiffening of such composites as follows. If we ignore surface tension effects, fluid-filled spherical inclusions in an elastic matrix will soften the composite, because the fluid inside the inclusions does not oppose any (static) shape changes. Conversely, for either large surface tension or small inclusions a significant energy penalty occurs for deformations of the unperturbed spherical inclusion shape. For sufficiently small liquid inclusions in soft materials, the energy balance is dominated by interfacial contributions.  Therefore, the tendency for inclusions to maintain sphericity increases relative to the case in which the same volume consisted of the matrix material itself (with no surface tension), thereby resulting in the composite being stiffer than the host elastic matrix alone. 

Here we generalize our treatment of soft composites with liquid inclusions \cite{MSW3phase, MSWa} to the case wherein the inclusions are elastic solids. In particular, we 
consider incompressible, homogeneous, isotropic linear-elastic spherical inclusions embedded in a homogenous, isotropic linear-elastic host matrix.  We treat the deformation of inclusions and their mechanical interaction with the surrounding host matrix, including the inclusion/host surface tension.   As we did with liquid inclusions \cite{MSW3phase, MSWa}, here we account for a finite volume fraction $\phi$ of inclusions through a three-phase generalized self-consistent (GSC) model \cite{Christensen79}, thereby allowing us to recover the dilute limit.  The effect of the inclusion/host surface tension is captured using a generalized Young-Laplace stress boundary condition, and the effective shear modulus of the composite is determined by imposing an energetic self-consistency condition. For brevity of exposition, we focus most of the discussion on the case of an incompressible matrix, and we examine the volume fraction-dependence of the effective elastic modulus as a function of the elastocapillary  $\mathbfup{\gamma^\prime}=L/R$ and the stiffness contrast $C$ parameters. We find that combining a local measurement, such as the effective inclusion strain, and a global measurement, such as the composite stiffness, is sufficient to simultaneously determine the surface tension and the inclusion stiffness. Finally, we discuss the case of an auxetic matrix in the context of the cloaking condition.

\section{The model}
\begin{figure}[htp]
\centering 
\includegraphics[trim={3cm 1cm 9cm 1cm}, width=1\columnwidth]{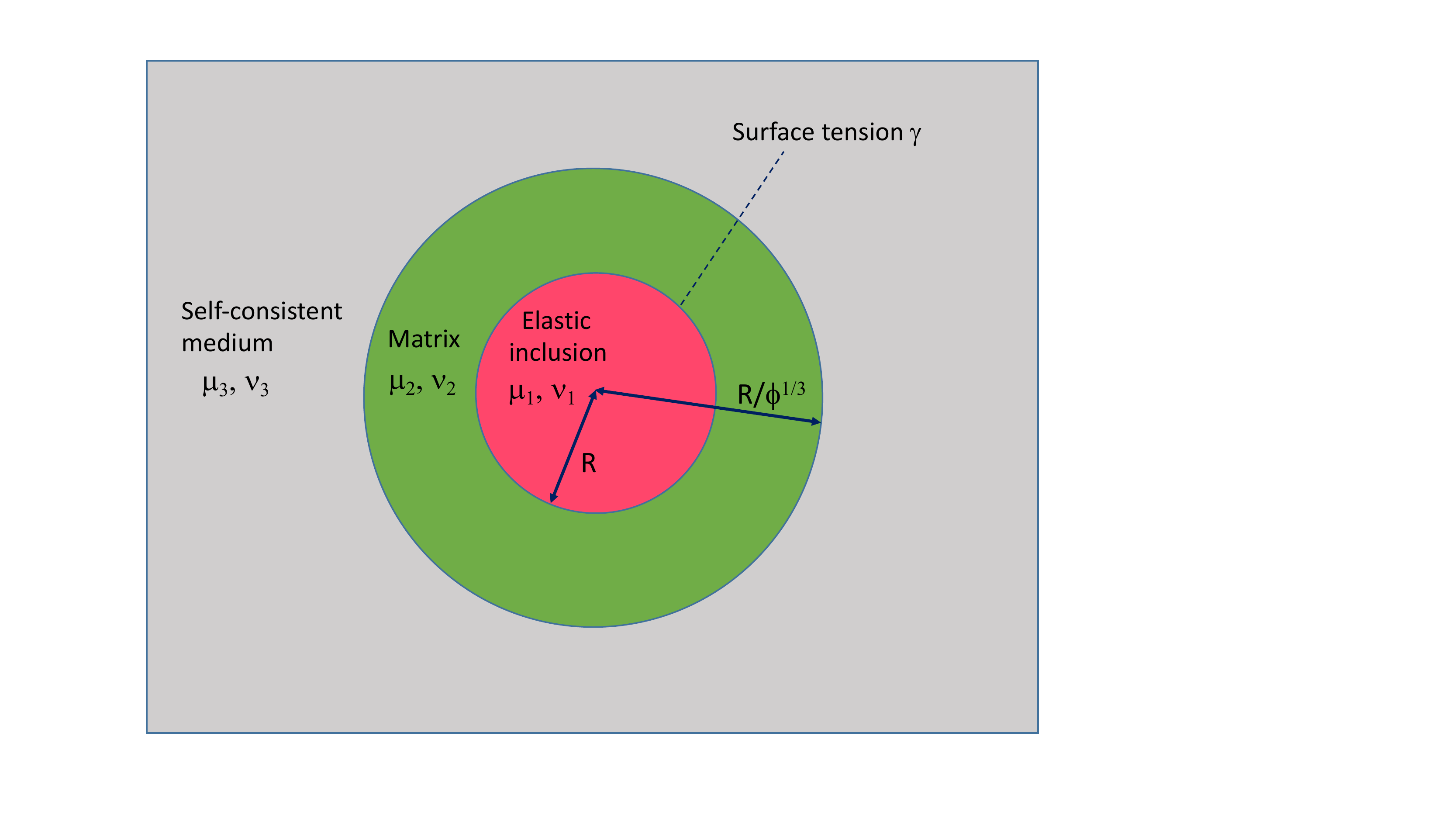}
\caption{Schematic of the model system under consideration. Each of the three components consists of a homogeneous isotropic linear-elastic material. Identical solid inclusions having elastic constants $\mu_1,\nu_1$ are embedded in a solid elastic matrix ($\mu_2,\nu_2$). The surrounding composite ($\mu_3,\nu_3$) is treated as an elastic medium with properties that are determined in a self-consistent manner. External body forces are ignored.}
\label{figcomposite}
\end{figure}

Consider a composite material with many identical, incompressible, homogeneous and isotropic linear-elastic spherical inclusions randomly embedded in a homogeneous and isotropic linear-elastic solid. 
The shear modulus and Poisson's ratio are $\mu_1, \nu_1$ for the inclusions, and $\mu_2, \nu_2$ for the matrix material, as shown in Fig.\,\ref{figcomposite}.
As we did in the case of  liquid inclusions \cite{MSW3phase, MSWa}, we extend the framework of the three-phase generalized self-consistent (GSC) theory of Kerner (see Ref.\,\cite{Kerner56}, and footnote 10 of Ref.\,\cite{MSW3phase}) to examine how surface tension at inclusion/matrix interfaces affects the composite's elastic moduli for arbitrary inclusion volume fraction $\phi$. 

The GSC approach treats the composite matrix as a \textit{single composite sphere} surrounded by an infinite medium of unknown effective elastic moduli $\mu_3, \nu_3$. The composite sphere consists of an elastic solid inclusion of radius $R$, surrounded by a concentric spherical shell of matrix material of radius $R/\phi^{1/3}$, thereby preserving the volume fraction $\phi$ of the original multi-inclusion system.  As described below, the elastic moduli $\mu_3, \nu_3$ of the region outside the composite sphere are determined by imposing an energetic self-consistency condition \cite{Christensen79}.\\

Now we provide expressions for the displacement fields $u_r^{(i)}, u_\theta^{(i)}$ and the linear-in-strain contributions $\sigma_{rr}^{(i)}, \sigma_{r\theta}^{(i)}$ to the stress fields, corresponding to a prescribed, purely deviatoric, far-field strain configuration, where the superscript ``$(i)$" refers to the inclusion ($i=1$), the matrix ($i=2$) or the composite effective medium ($i=3$) phase. The origin of a spherical polar coordinate system, ($r, \theta, \varphi$), is placed at the center of the composite sphere.  The following far-field ($r \rightarrow \infty$) displacements are imposed; 
\be 
u_r^0=2\varepsilon_{A}^0 r\, \mathcal{P}_2(\cos \theta), \quad  
u_\theta^0=\varepsilon_{A}^0 r\, \frac{d \mathcal{P}_2(\cos \theta)}{d \theta}, \quad 
u_\varphi^0=0\;,
\label{remote displacements}
\ee 
where $\mathcal{P}_2$ is the Legendre polynomial of order 2.  The corresponding deviatoric far-field strain configuration is expressed as
\be
\varepsilon_{xx}^0= \varepsilon_{yy}^0=-\varepsilon_{A}^0, \quad\quad \varepsilon_{zz}^0=2 \varepsilon_{A}^0\;. 
\label{deviatoricstrain}
\ee
Based on the azimuthal ($z$-axis) symmetry of the strained system, the displacement fields $u_r^{(i)}, u_\theta^{(i)}$ and the linear-in-strain contributions $\sigma_{rr}^{(i)}, \sigma_{r\theta}^{(i)}$ to the stress fields take the form \cite[e.g.,][]{Lure64, Duan05, Stylesoft15}; 
\begin{multline}
u_r^{(i)}(\rho,\theta)=\left(\mathcal{F}_i+\frac{\mathcal{G}_i}{\rho^3}\right)r+\mathcal{P}_2(\cos \theta)\\
\times\left[12 \nu_i \mathcal{A}_i \rho^2+2\mathcal{B}_i+2\frac{(5-4\nu_i)\mathcal{C}_i}{\rho^3}-3\frac{\mathcal{D}_i}{\rho^5}\right]r\;,
\label{radialansatz}
\end{multline}
\begin{multline}
u_\theta^{(i)}(\rho,\theta)=\frac{d\mathcal{P}_2(\cos \theta)}{d\theta}\\ \times \left[(7-4\nu_i)\mathcal{A}_i \rho^2+\mathcal{B}_i+2\frac{(1-2\nu_i)\mathcal{C}_i}{\rho^3}+\frac{\mathcal{D}_i}{\rho^5}\right]r\;,
\label{polaransatz}
\end{multline}
\begin{multline}
\sigma_{rr}^{(i)}(\rho,\theta)=2\mu_i 
\left\{-2 \frac{\mathcal{G}_i}{\rho^3}+\frac{\mathcal{F}_i (1+\nu_i)}{1-2\nu_i}+\right.\\
\left.\mathcal{P}_2(\cos \theta)\left[-6\nu_i\mathcal{A}_i \rho^2+2\mathcal{B}_i-\frac{4(5-\nu_i)}{\rho^3}\mathcal{C}_i+\frac{12 \mathcal{D}_i}{\rho^5} \right]\right\}\,,
\label{radialstress}
\end{multline}
and
\begin{multline}
\sigma_{r\theta}^{(i)}(\rho,\theta)=2\mu_i \frac{d \mathcal{P}_2(\cos \theta)}{d\theta} \\
\times\left[(7+2\nu_i)\mathcal{A}_i \rho^2+\mathcal{B}_i +\frac{2(1+\nu_i)}{\rho^3}\mathcal{C}_i  -\frac{4 \mathcal{D}_i}{\rho^5} \right]\,,
\label{polarstress}
\end{multline}
where $\rho\equiv r/R$ is the radial coordinate scaled by the inclusion radius, and the 18 coefficients $\mathcal{A}_i$ through $\mathcal{G}_i$ are determined from the boundary and regularity conditions that we discuss next.

In the inclusion region, the linear-in-strain stress tensor $\sigma^{(1)}$ is supplemented by a strain-independent, hydrostatic stress tensor $\sigma^{*(1)}$, referenced to the surface tension at the inclusion interface and present in the stress-free configuration viz., 
\be 
\sigma_{rr}^{*(1)}\vert_{r=R}=\sigma_{\theta\theta}^{*(1)}\vert_{r=R}=-2\gamma/R, \quad\quad \sigma_{r\theta}^{*(1)}\vert_{r=R}=0\,.
\label{hydrocomponents}
\ee

Combining the stress-displacement relationship with Eq.\,(\ref{remote displacements}) yields the far-field stresses as 
\be
\sigma_{rr}^0=4 \varepsilon_A^0 \mu_3 \mathcal{P}_2(\cos \theta)\,,   \quad 
\sigma_{r\theta}^0=2 \varepsilon_A^0 \mu_3 \frac{d \mathcal{P}_2(\cos \theta)}{d \theta}\,.
\label{farsigma}
\ee
We stress that within the matrix ($i=2$) and the effective medium ($i=3$) regions, the overall stress is linear-in-strain, and hence the components are simply expressed by Eqs.\,(\ref{radialstress}) and (\ref{polarstress}).
The additional information provided by Eqs.\,(\ref{hydrocomponents}) and (\ref{farsigma}) closes the problem, providing the complete set of conditions on the 18 coefficients $\mathcal{A}_i$ -- $\mathcal{G}_i$ associated with the total stress field as described presently.\\

Six of the 18 coefficients in Eqs.\,(\ref{radialansatz}-\ref{polarstress}) are determined by the far-field stress/strain and regularity at the origin, viz., $\mathcal{A}_3=\mathcal{F}_3=\mathcal{C}_1=\mathcal{D}_1=\mathcal{G}_1=0$, and $\mathcal{B}_3=\varepsilon_A^0$. In addition, twelve independent conditions can be written for the remaining twelve unknowns $\mathcal{A}_1, \mathcal{B}_1, \mathcal{F}_1,\mathcal{A}_2, \mathcal{B}_2, \mathcal{C}_2, \mathcal{D}_2, \mathcal{F}_2, \mathcal{G}_2, \mathcal{C}_3, \mathcal{D}_3, \mathcal{G}_3$: (i) Six of these conditions arise from continuity of displacement and stress at the composite sphere surface ($\rho=\alpha \equiv 1/ \phi^{1/3}$), (ii) three from continuity of displacement at the inclusion boundary, and (iii) three from the stress boundary conditions associated with the generalized Young-Laplace condition.  In detail, at the composite sphere surface $\rho=\alpha$ the continuity of displacement for $u_\theta$ ($u_r$) provides one (two) condition(s), and that for stress $\sigma_{r\theta}$ ($\sigma_{rr}$) provides one (two) condition(s), for a total of six conditions arising from the four continuity equations which are
\begin{align}
u_r^{(2)}(\alpha, \theta)&=u_r^{(3)}(\alpha, \theta),\quad u_\theta^{(2)}(\alpha, \theta)=u_\theta^{(3)}(\alpha, \theta) \qquad \textrm{and} \nonumber \\
\sigma_{rr}^{(2)}(\alpha, \theta)&=\sigma_{rr}^{(3)}(\alpha, \theta),\quad \sigma_{r\theta}^{(2)}(\alpha, \theta)=\sigma_{r\theta}^{(3)}(\alpha, \theta).
\label{continuityexternalur}
\end{align}
Similarly, continuity of displacement for  $u_\theta$ ($u_r$) provides one (two) condition(s) at the inclusion boundary, for a total of three independent conditions expressed via the two (polar and radial) constraints as
\be
u_r^{(1)}(1, \theta)=u_r^{(2)}(1, \theta),\quad u_\theta^{(1)}(1, \theta)=u_\theta^{(2)}(1, \theta)\;.
\ee
The final three constraints arise from the stress boundary conditions at the inclusion surface, treated using the generalized Young-Laplace equation; 
\be
\sigma^{(2)} \cdot \textbf{n}-(\sigma^{(1)}+\sigma^{*(1)}) \cdot \textbf{n}=\gamma \mathcal{K} \textbf{n}\,,
\label{YoungLaplace}
\ee
where $\sigma^{(2)}$ denotes the stress tensor in the matrix region, $\sigma^{(1)}$ and $\sigma^{*(1)}$ denote the linear-in-strain and the hydrostatic components of the stress tensor in the inclusion region respectively.  Here, \textbf{n} is the outward normal to the inclusion surface, $\mathcal{K}$ is its total curvature, and the surface tension is $\gamma$, as discussed in the introduction. To leading order, both $\textbf{n}$ and $\mathcal{K}$ can be expressed in terms of the displacement field $\textbf{u}$ \cite[see Eqs.\,(24) and (28) of Ref.][]{Stylesoft15}, 
%
and as a result equation (\ref{YoungLaplace}) can be rewritten as
\begin{multline}
\nonumber
\left[\begin{array}{c}\sigma_{rr}+\sigma_{r\theta}\,(u_\theta-\frac{\partial u_r}{\partial \theta})R^{-1} \\ \sigma_{\theta r}+\sigma_{\theta\theta}\,(u_\theta-\frac{\partial u_r}{\partial \theta})R^{-1}\end{array}\right]^{(2)}_{(1)}-   
\left[\begin{array}{c}-2\gamma/R \\ 
-2\gamma(u_\theta-\frac{\partial u_r}{\partial \theta})/R^2\end{array}\right]^{(1)}=\\
= \frac{\gamma}{R^2}\left[\begin{array}{c} 2R-(2 u_r+ \cot \theta \, \frac{\partial u_r}{\partial \theta}+\frac{\partial^2 u_r}{\partial \theta^2}) \\ 2(u_\theta-\frac{\partial u_r}{\partial \theta})\end{array}\right]\;. 
\end{multline}
Therefore, at first order in $\textbf{u}$ we have
\be
\left[\begin{array}{c}\sigma_{rr} \\ \sigma_{\theta r}\end{array}\right]^{(2)}_{(1)}= \left[\begin{array}{c} -\frac{\gamma}{R^2}(2 u_r+ \cot \theta \, \frac{\partial u_r}{\partial \theta}+\frac{\partial^2 u_r}{\partial \theta^2}) \\ 0\end{array}\right]\;, 
\label{vectoryoung}
\ee
where $\left[\right]^{(2)}_{(1)}$ denotes the jump in the stresses $\sigma_{rr}$ and $\sigma_{r\theta}$ across the matrix/inclusion (2/1) interface. Thus, the twelve unknowns $\mathcal{A}_1$,$\mathcal{B}_1$,$\mathcal{F}_1$,$\mathcal{A}_2$,$\mathcal{B}_2$,$\mathcal{C}_2$,$\mathcal{D}_2$,$\mathcal{F}_2$,$\mathcal{G}_2$,$\mathcal{C}_3$,$\mathcal{D}_3$,$\mathcal{G}_3$ are determined by the twelve equations (\ref{continuityexternalur})-(\ref{vectoryoung}) in terms of the unknown parameters $\mu_3$, $\nu_3$.  Finally, the effective shear modulus $\mu_3$ is determined by imposing the following \emph{energetic self-consistency condition} \cite{Christensen79}: the total mechanical work $W$ associated with the presence of the composite sphere inside the infinite effective medium vanishes; 
\be W=0\,. \label{energeticcondition} \ee
%
We use Eshelby's formula to evaluate the total strain energy \cite[Eq. 5.1 of][]{Eshelby56}, \cite[see also \S 4.1 of][]{Eshelby56} to find \cite{Stylesoft15} 
\begin{align}
W = & \frac{1}{2}\int_{S^{ext (+)}}[n_i \sigma_{ij}^0 u_j-n_i \sigma_{ij} u_j^0]dS \nonumber \\
&-\frac{\gamma}{2}\int_{S^{int}} \mathcal{K} u\cdot n\, dS+\gamma \Delta S, 
\label{workcompositeinclusion}
\end{align}
where $S^{int}$ refers to the inclusion/matrix interface, $S^{ext (+)}$ refers to the exterior of the composite sphere/external effective medium interface, and $ \Delta S$ denotes the change of the interfacial area associated with the deformation. The last two summands of Eq. (\ref{workcompositeinclusion}) cancel exactly whenever the deformation of the inclusion is volume-preserving \cite{MSW3phase}, which is the case for purely deviatoric far-field strain conditions (\ref{deviatoricstrain}). 

\begin{figure*}[t]
\centering 
\begin{tabular}{ccc}
\includegraphics[trim={4cm 5cm 0 5cm}, width=0.7\columnwidth]{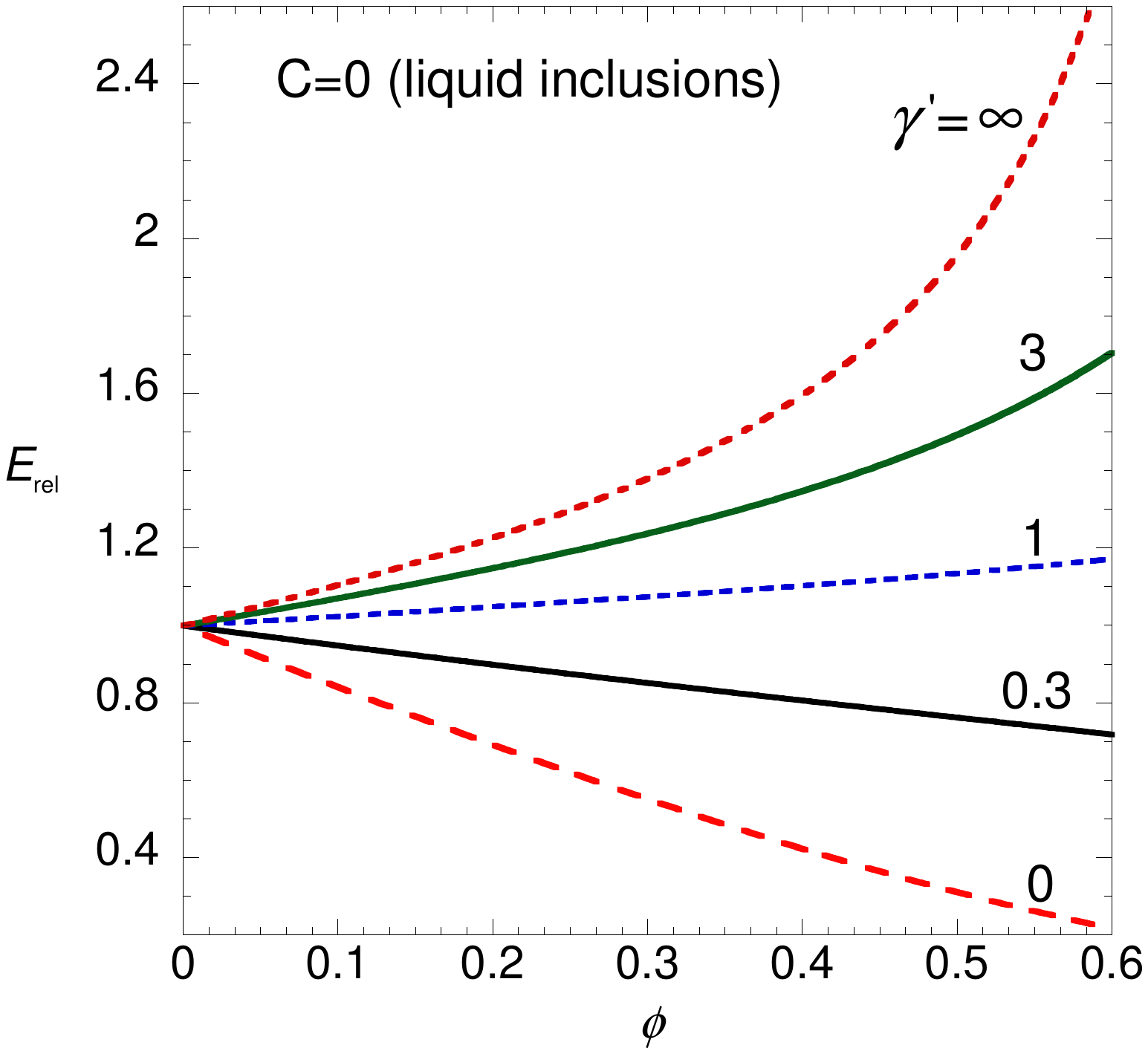}&
\includegraphics[trim={4cm 5cm 0 5cm}, width=0.7\columnwidth]{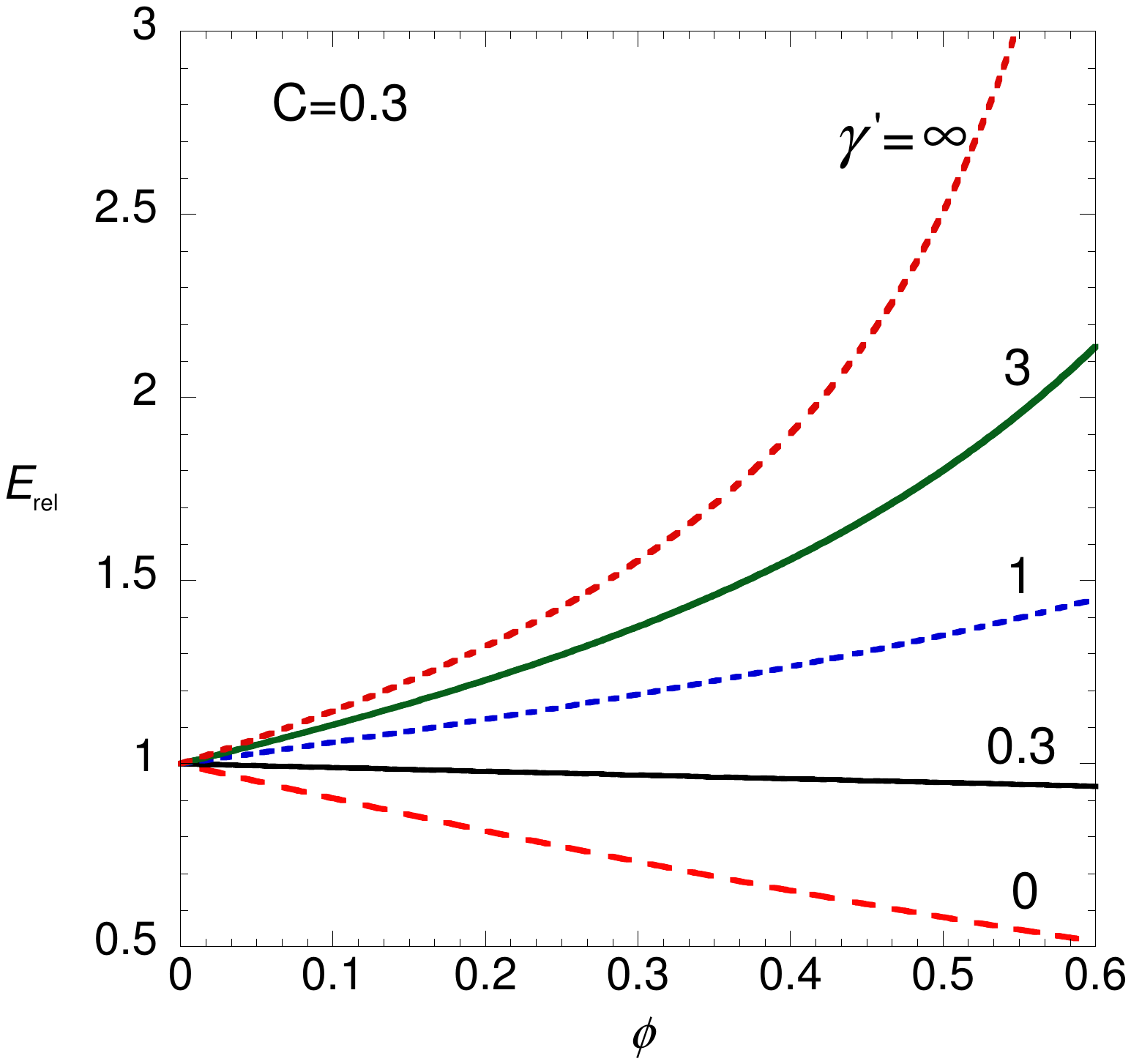}&
\includegraphics[trim={4cm 5cm 0 5cm}, width=0.7\columnwidth]{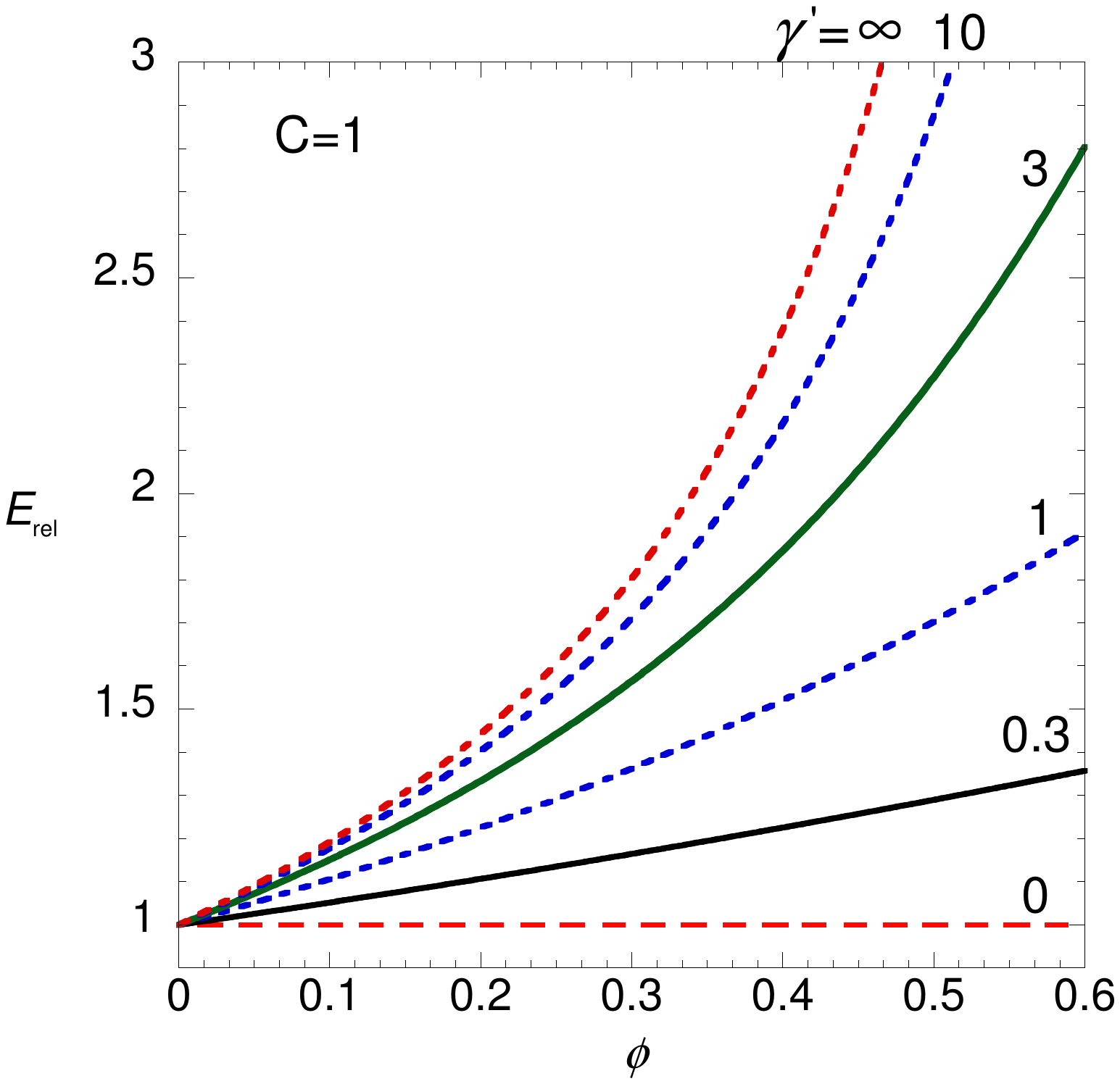}\\
\includegraphics[trim={4cm 8cm 0 9cm}, width=0.7\columnwidth]{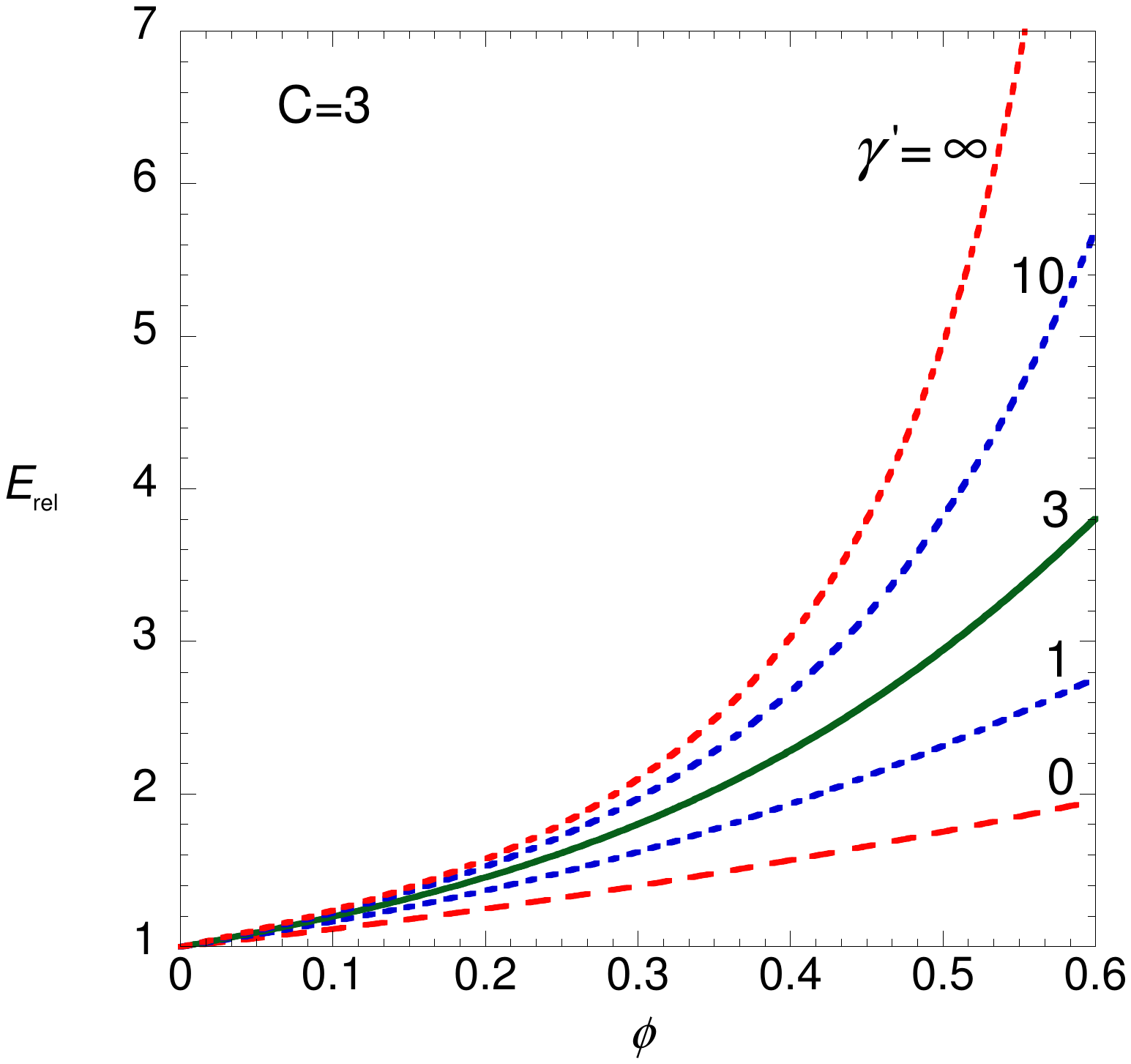}&
\includegraphics[trim={4cm 8cm 0 9cm}, width=0.7\columnwidth]{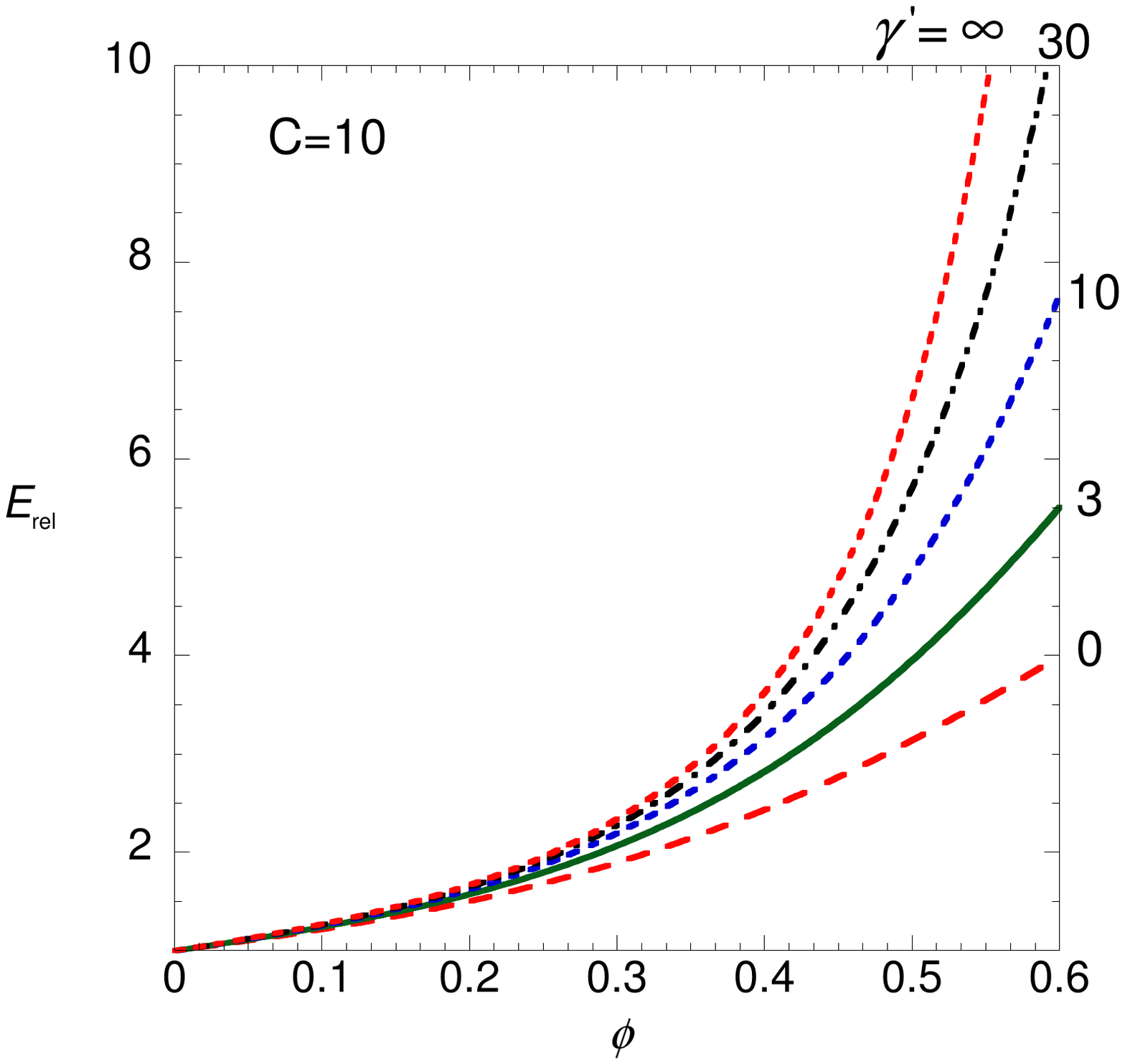}&
\includegraphics[trim={4cm 8cm 0 9cm}, width=0.7\columnwidth]{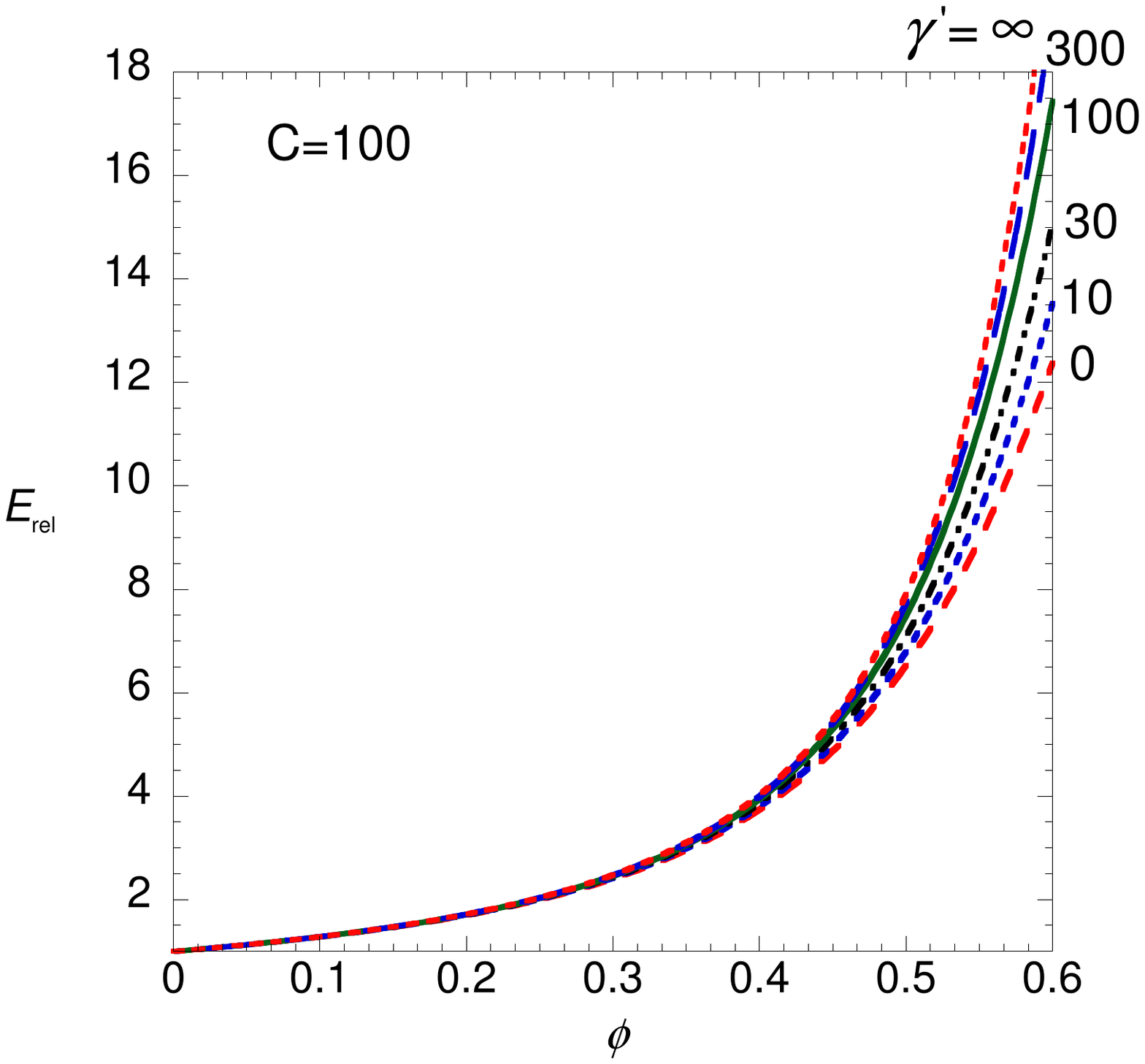}
\end{tabular}
\caption{$E_{rel}$ versus $\phi$ in the case of an incompressible matrix for the entire range of $\mathbfup{\gamma^\prime}$, from the surface-free regime $\mathbfup{\gamma^\prime}=0$ to the surface tension-dominated regime $\mathbfup{\gamma^\prime}=\infty$. In each panel, the bottom red long-dashed curve for $\mathbfup{\gamma^\prime}=0$ corresponds to the standard Eshelby-like theory. In the top row $C=0\text{ (liquid inclusions)}, 0.3,1$ and in the bottom row $C= 3$, $10$, and $100$. Note that the vertical axes have different scales on each panel to visualize the relevant range of behavior. Importantly, the interfacial stress has appreciable effects on the effective Young's modulus even for inclusions a hundred times more stiff than their host matrix.  The softening of the composite occurs for $\mathbfup{\gamma^\prime} < \mathbfup{\gamma^\prime}_{cl}(C)\equiv (1-C)(19 C+16)/[12(2+5C)]$, and  the stiffening of the composite occurs for $\mathbfup{\gamma^\prime} > \mathbfup{\gamma^\prime}_{cl}(C)$, with both softening and stiffening possible when $C<1$.  However, as expected, if $C>1$ the inclusions stiffen the matrix for any value of $\mathbfup{\gamma^\prime}$. Exact mechanical cloaking of the inclusions (i.e. $E_{rel}=1$) is found at $\mathbfup{\gamma^\prime}=\mathbfup{\gamma^\prime}_{cl}(C)$ whenever $C \leq 1$.}
\label{figseveralC} 
\end{figure*}

For incompressible inclusions, the self-consistency condition of Eq. (\ref{energeticcondition}) reduces to a quadratic equation for the relative effective shear modulus $\mu_{rel}\equiv \mu_3/\mu_2$. We have found that \cite{MSW3phase}
\begin{align}
W&=\frac{1}{2}\int_{S^{ext (+)}}[\sigma_{rr}^0 u_r+\sigma_{r\theta}^0 u_\theta - \sigma_{rr} u_r^0 - \sigma_{r\theta} u_\theta^0]dS \nonumber \\
&=-\frac{48 \pi \mu_3 R^3 \varepsilon_A^0 (\nu_3-1)}{\alpha^2}\, \mathcal{C}_3\;, 
\label{workformula}
\end{align}
and hence Eqs.\,(\ref{energeticcondition}) and (\ref{workformula}) imply that $\mathcal{C}_3=0$ which, combined with the solution of the system of equations (\ref{continuityexternalur})-(\ref{vectoryoung}), yields the following quadratic condition on $\mu_{rel}$:
\be
2 R\mu_2(a_0+a_1 \mu_{rel}+a_2 \mu_{rel}^2)+ \gamma (b_0+b_1 \mu_{rel}+b_2 \mu_{rel}^2)=0\,,
\label{quadraticgeneral}
\ee
where the coefficients $a_0,a_1,a_2,b_0,b_1,b_2$ depend on $\phi$, $\nu_2$, and $C\equiv \mu_1/\mu_2$, as shown in Appendix \ref{AppendixA} for the incompressible matrix case ($\nu_2=1/2$). Hence, for incompressible inclusions, $\mu_{rel}$ is a function of the 4 parameters $\phi$, $\nu_2$, $C$ and ${\gamma}/{ (R \mu_2)}$.
If we define the elastocapillary length as $L\equiv \gamma/E_2$, based on the matrix phase of the composite sphere, then $\mathbfup{\gamma^\prime}\equiv L/R=\gamma/(E_2 R)$ is a key dimensionless parameter.  
In the limit of small/large inclusions (large/small $\mathbfup{\gamma^\prime}$), Eq.\,(\ref{quadraticgeneral}) simplifies to $b_0+b_1 \mu_{rel,R \ll L}+b_2 \mu_{rel,R \ll L}^2=0$.
Thus, for each $\phi$ the solution of this equation, $\mu_{rel,R \ll L}[\phi,\nu_2,C]$, gives the upper limit of rigidity among all $\mathbfup{\gamma^\prime}$-curves.  In the limit $\phi\rightarrow 0$, for an incompressible matrix and very large $\mathbfup{\gamma^\prime}$, the shear and Young's relative moduli are $ \sim 1+\phi(2+5C)/[2(1+C)]$ (see the $\mathbfup{\gamma^\prime}=\infty$ curves in Fig.\,\ref{figseveralC}).
Finally, this dilute-case relationship displays an expected monotonic dependence on the stiffness contrast parameter $C$, and is consistent with the behavior found for both liquid inclusions ($C=0$) \cite{MSW3phase}, and rigid inclusions ($C \rightarrow \infty$) \cite{Eshelby57}. 

\section{Composite Properties of Incompressible Hosts ($\nu_2=1/2$)}

\subsection{Effective Medium Properties and Cloaking} \label{subcloaking}

We consider the case of an incompressible host matrix, in which $E_{rel}\equiv\left(\frac{E_3}{E_2}\right)=\left(\frac{\mu_3}{\mu_2}\right)\equiv\mu_{rel}$ and $\nu_2=1/2$.  Now, by making use of Eqs.\,(\ref{dropshaper}) and (\ref{dropshapetheta}), we treat the deformation of the inclusion phase and evaluate the effective inclusion strain $\varepsilon_{inc}\equiv(l-2R)/R=2 u_r(1,0)/R$, where $l$ denotes the major axis of the inclusion \footnote{Here we define $\varepsilon_{inc}=(l-2R)/R$ rather than $\varepsilon_{inc}=(l-2R)/2R$, to make clear the comparison with $\varepsilon_d=\varepsilon_{inc}\vert_{C=0}$ considered in \cite{Stylesoft15} and \cite{MSW3phase} for liquid inclusions.}.  In terms of $\alpha=\phi^{-1/3}$, $\mathbfup{\gamma^\prime}$, and the solution of Eq.(\ref{quadraticgeneral}), $\mu_{rel}=\mu_3/\mu_2$, the radial and polar displacements of the inclusion surface are
\be
\frac{u_r(1,\theta)}{R}=100 \alpha ^3 \mu _{\text{rel}} \varepsilon ^o{}_A \; \frac{f_1}{f_2+\mathbfup{\gamma^\prime} f_3}\;\mathcal{P}_2(\cos \theta)
\label{radshape}
\ee
and
\be 
\frac{u_\theta(1,\theta)}{R}=25 \mu_{rel} \alpha^3\,\varepsilon_A^0 \;\frac{f_4+\mathbfup{\gamma^\prime} f_5}{f_2+\mathbfup{\gamma^\prime} f_3}\;\frac{d \mathcal{P}_2(\cos \theta)}{d \theta}\,,
\label{thetashape}
\ee
respectively, where the coefficients $f_1$--$f_5$ are in Appendix \ref{AppendixB}.
When $R \ll L $ and thus $\mathbfup{\gamma^\prime} \gg 1$ then, whatever the stiffness of the inclusions, the radial displacement is very small and 
the inclusions remain nearly spherical. In the opposite limit wherein $R \gg L$ and thus $\mathbfup{\gamma^\prime} \ll 1$,
the inclusion shape is again scale-invariant, in agreement with the theory for the case of purely bulk elasticity, and the corresponding effective inclusion strain is
$\varepsilon_{inc}= 200 \mu_{rel} \alpha^3 \varepsilon_A^0 f_1/(f_2+\mathbfup{\gamma^\prime} f_3)$.
In the dilute limit ($\phi\rightarrow 0$) of the incompressible case, the inclusion's effective strain and shape reduce to 
\be \varepsilon_{inc}\vert_{\phi\rightarrow 0}=\frac{20 \varepsilon_A^0}{2C+3+120 \mathbfup{\gamma^\prime} \frac{C+1}{19C+16}}\,,
\label{diluteinclusionstrain}
\ee
\be
\nonumber
\frac{u_r(1,\theta)}{R}\vert_{\phi\rightarrow 0}=\frac{10 \varepsilon_A^0 \mathcal{P}_2(\cos \theta)}{2C+3+120 \mathbfup{\gamma^\prime} \frac{C+1}{19C+16}}\,,
\ee
and
\be
\nonumber
\frac{u_\theta(1,\theta)}{R}\vert_{\phi\rightarrow 0}=\frac{5 \varepsilon_A^0(1+\frac{24}{19C+16} \mathbfup{\gamma^\prime})\frac{d \mathcal{P}_2(\cos \theta)}{d \theta}}{2C+3+120 \mathbfup{\gamma^\prime} \frac{C+1}{19C+16}} \;, 
\ee
in agreement with analogous results for liquid inclusions \cite{Stylesoft15, MSW3phase}.   Importantly, in the dilute limit at the exact cloaking point defined by
\be 
\mathbfup{\gamma^\prime}_{cl}(C)\equiv (1-C)(19 C+16)/[12(2+5C)]\,,
\label{cloakingpoint}
\ee 
the {\em inclusions} will {\em stretch less than} the {\em host} material viz., $(l-2R)/(2R)=10 (5 C+2) \varepsilon^0_A/(19 C+16)$. Indeed, the corresponding host material strain, $\varepsilon_{zz}^0=2 \varepsilon^0_A$, exceeds that of the inclusions whenever the cloaking condition occurs; the inclusions are softer than the matrix. However, for any $C<1$ within the softening regime, the {\em inclusions} will {\em stretch the same amount} as the {\em host} material when $\mathbfup{\gamma^\prime}=(1-C) (19 C+16)/[60 (1+C)] < \mathbfup{\gamma^\prime}_{cl}(C)$. When $\phi,\mathbfup{\gamma^\prime} \ll 1$ the predictions of Eshelby's theory \cite{Eshelby57}, and effective inclusion strain in the liquid case \cite{Stylesoft15}, $(10/3) \varepsilon_{zz}^\infty$, are recovered, whereas when $\mathbfup{\gamma^\prime} \gg 1$, for arbitrary values of $\phi$, we find a nearly unperturbed spherical inclusion shape.

\begin{figure*}[t]
\centering 
\begin{tabular}{ccc}
\includegraphics[trim={4cm 5cm 0 5cm}, width=0.68\columnwidth]{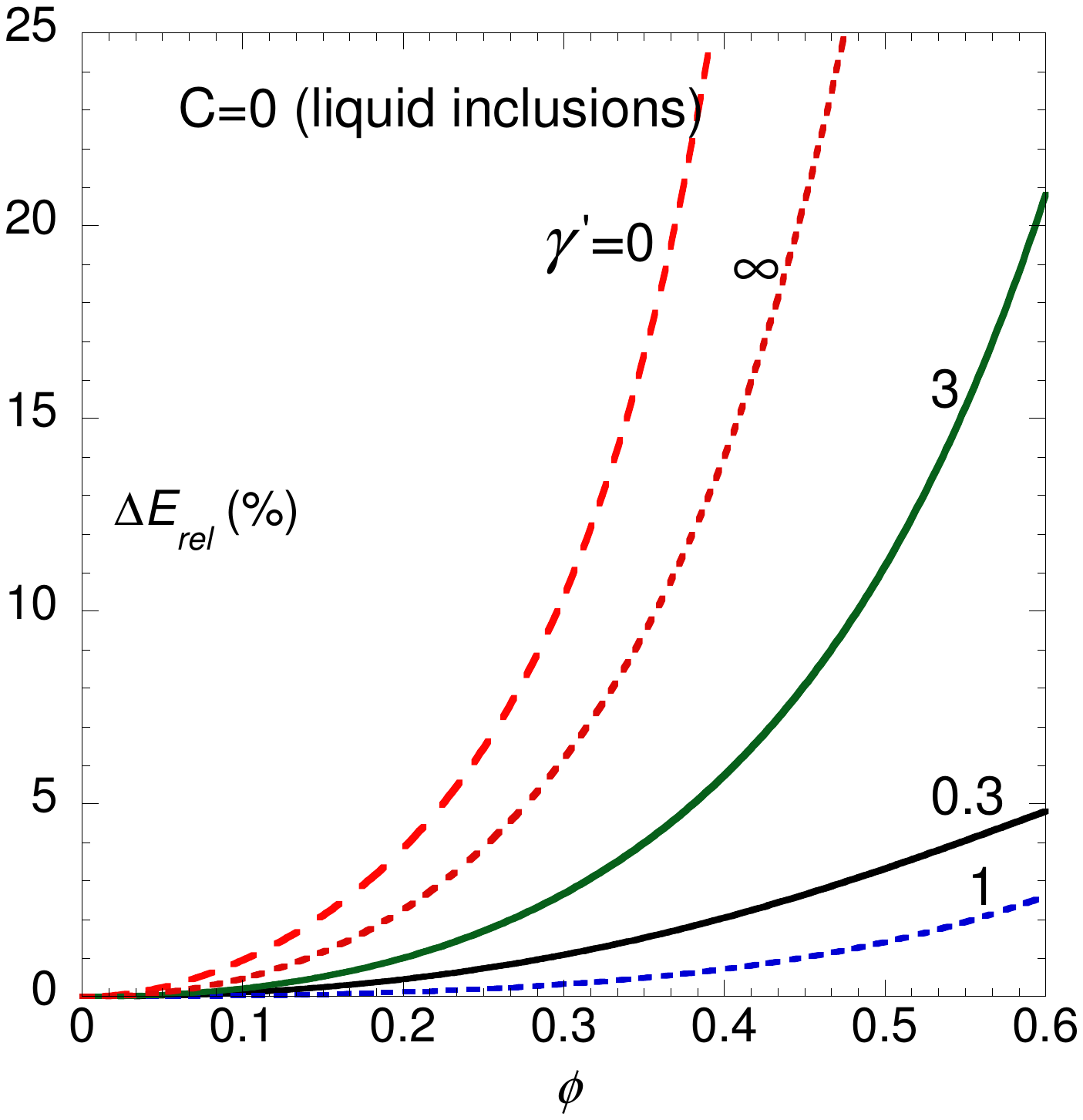}&
\includegraphics[trim={4cm 5cm 0 5cm}, width=0.68\columnwidth]{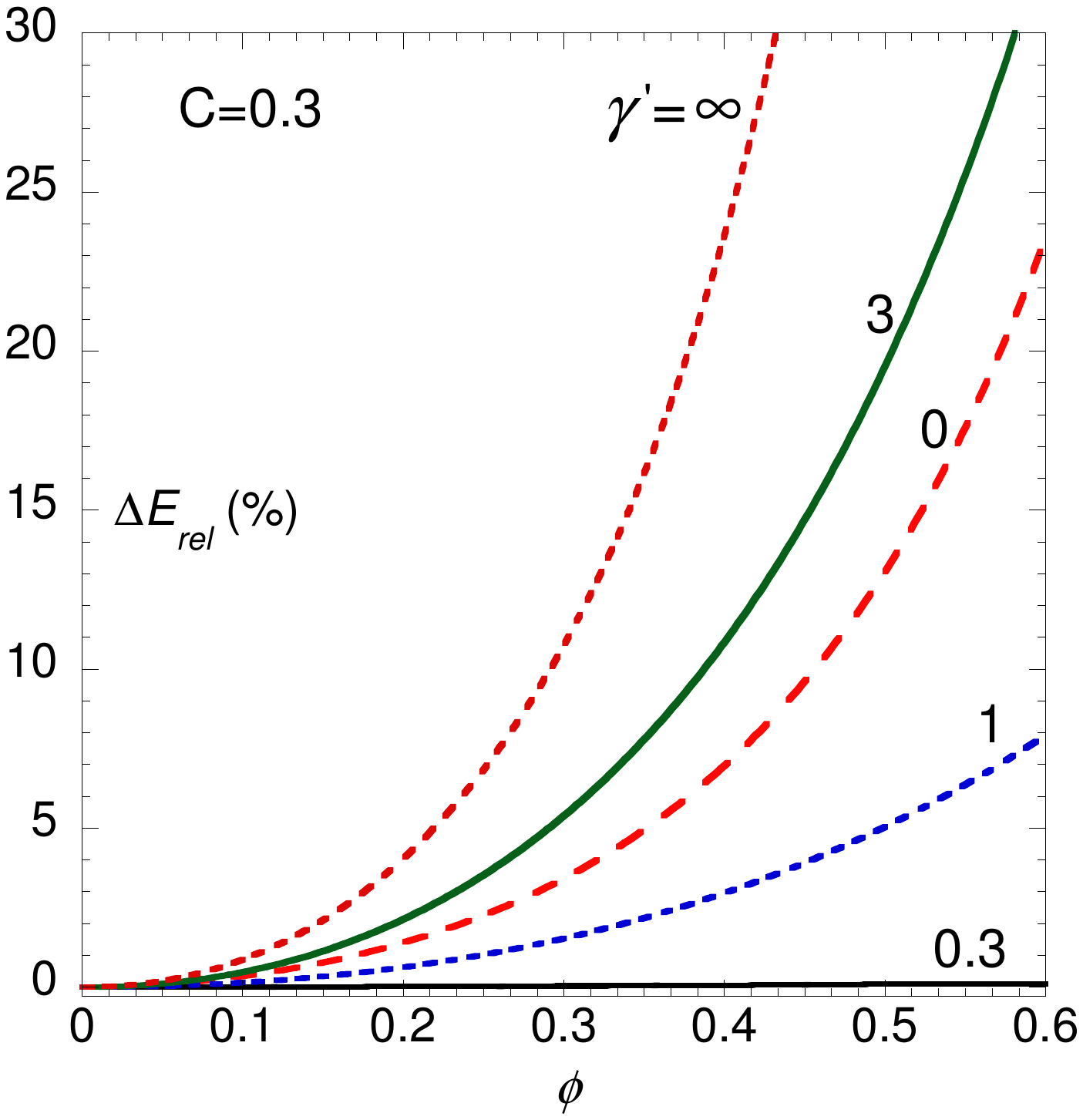}&
\includegraphics[trim={4cm 5cm 0 5cm}, width=0.68\columnwidth]{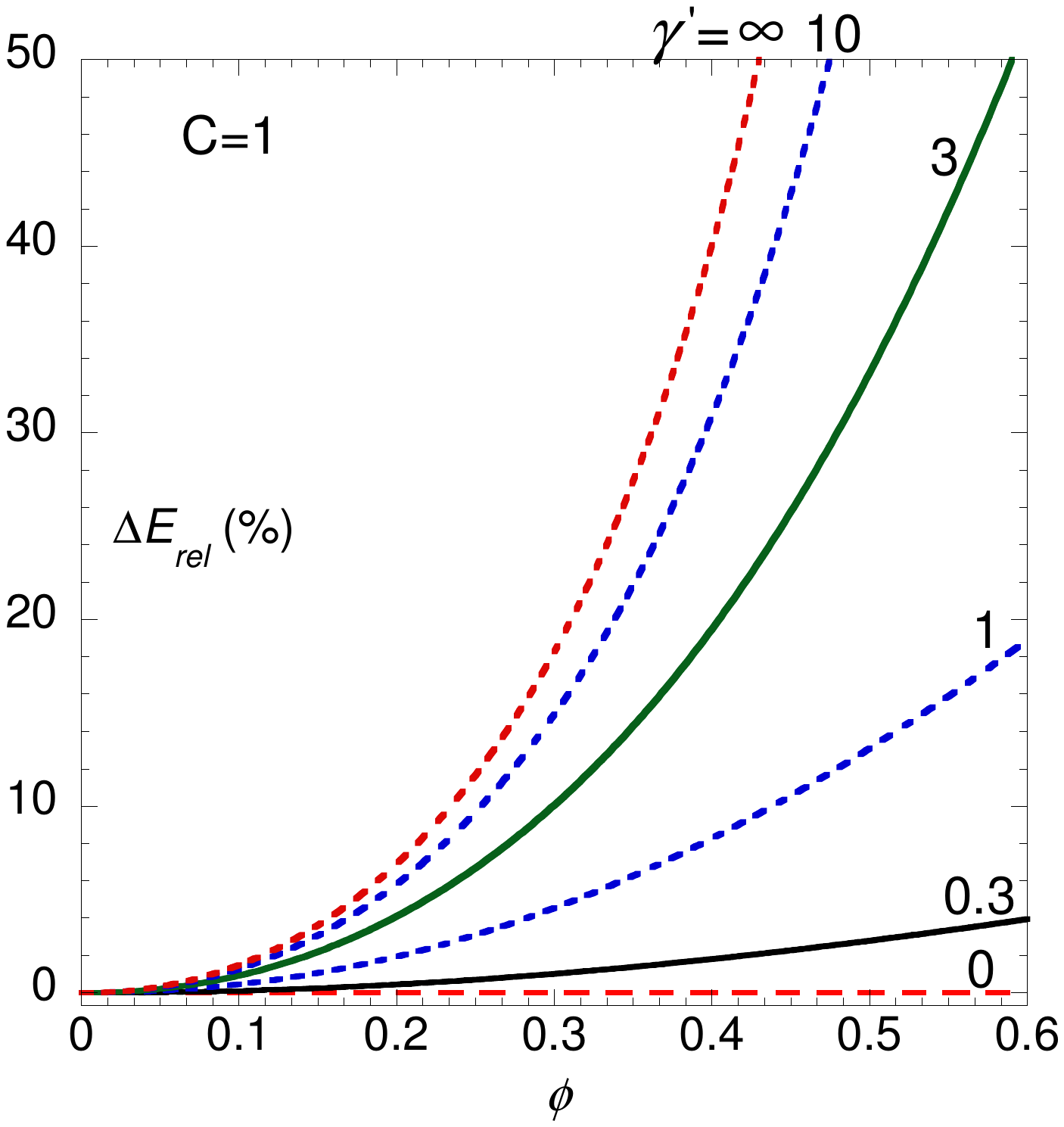}\\\\\\
\includegraphics[trim={4cm 8cm 0 9cm}, width=0.68\columnwidth]{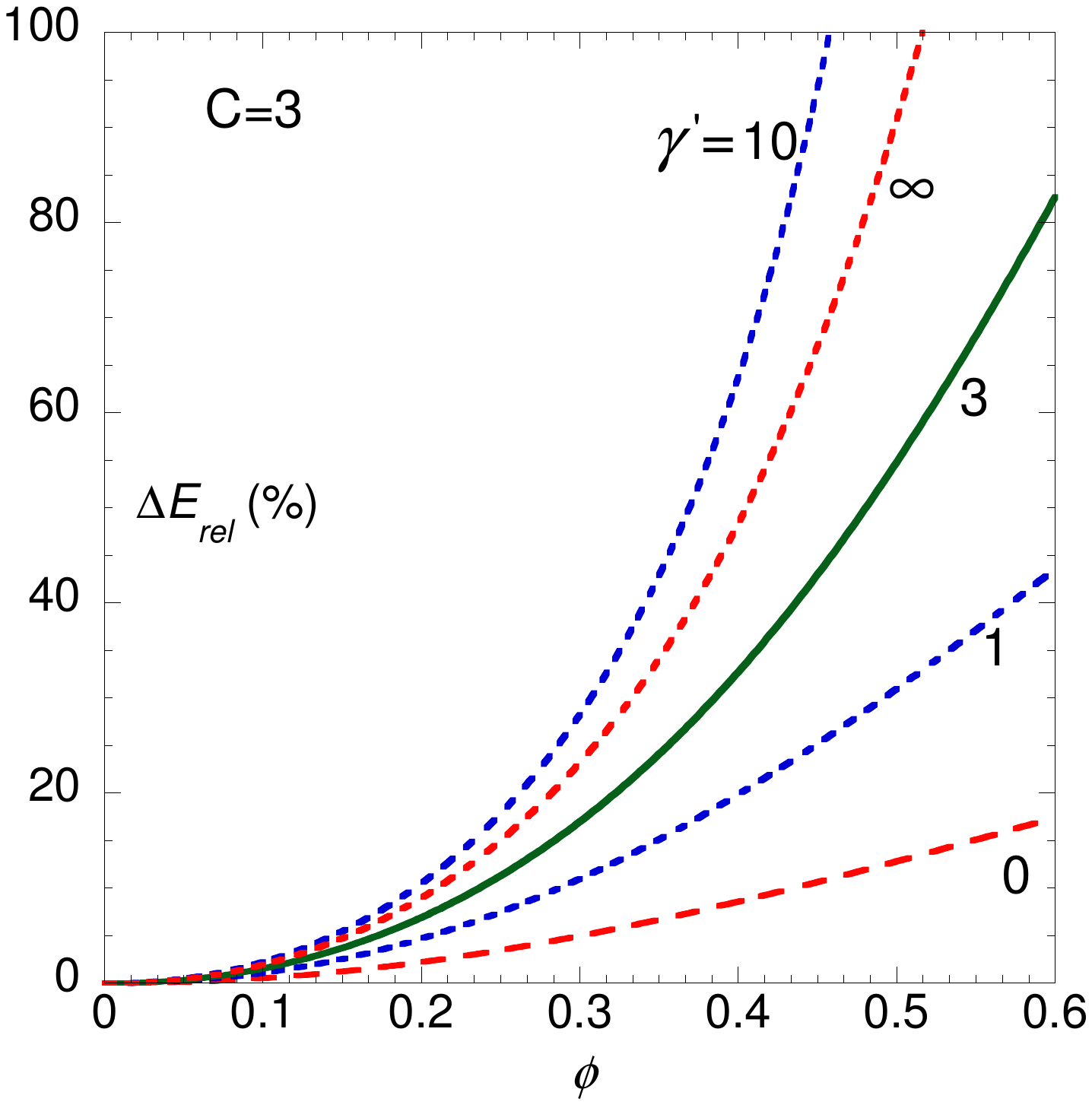}&
\includegraphics[trim={4cm 8cm 0 9cm}, width=0.68\columnwidth]{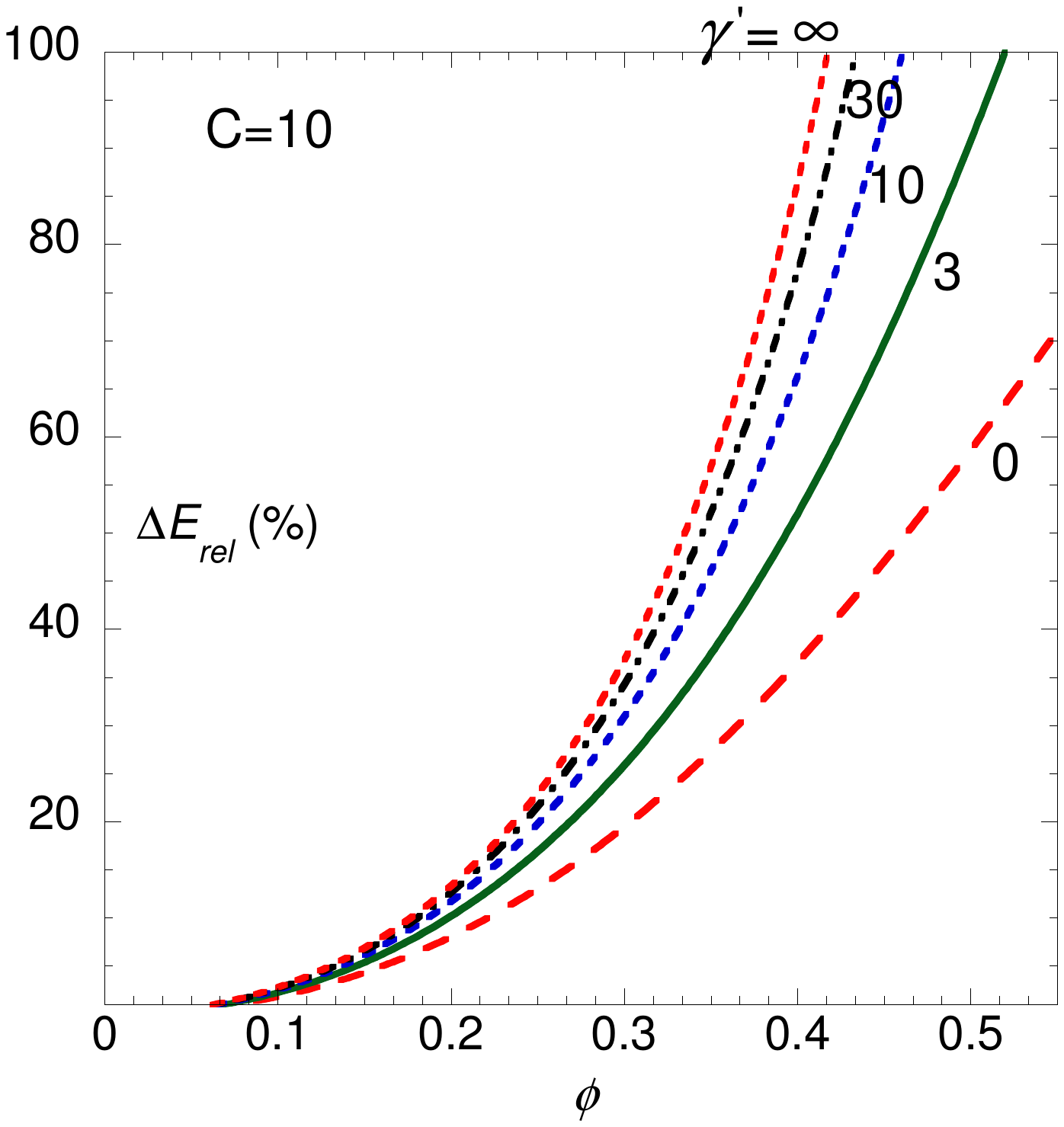}&
\includegraphics[trim={4cm 8cm 0 9cm}, width=0.68\columnwidth]{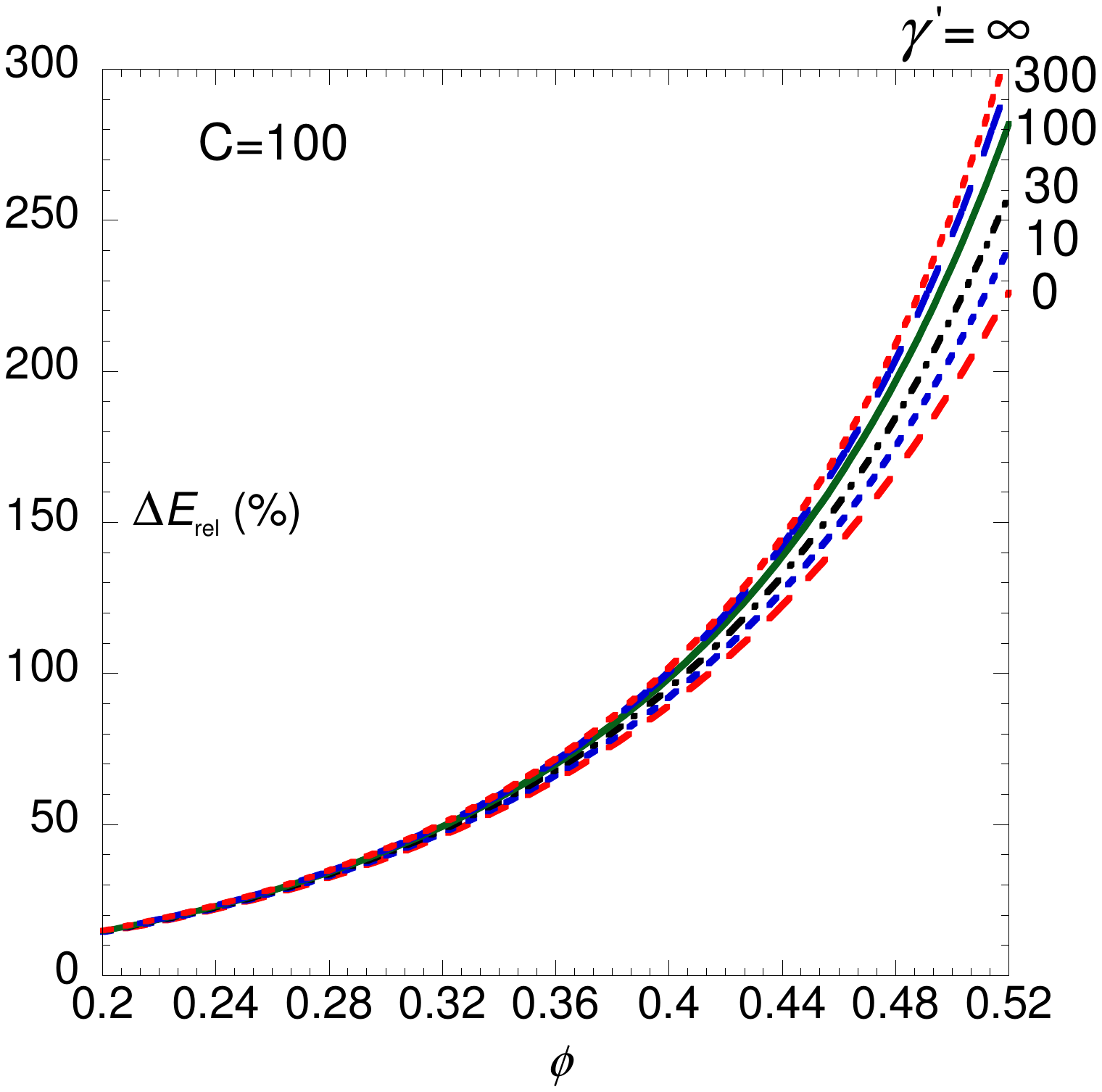}
\end{tabular}
\caption{For the case of an incompressible matrix, the percentage deviation $ \Delta E_{rel}(\%)$ from the linear-in-$\phi$ (dilute) approximation is plotted versus $\phi$ across the entire range of $\mathbfup{\gamma^\prime}$, from the bulk elasticity-dominated regime, for which $\mathbfup{\gamma^\prime}=0$, to the surface tension-dominated regime, for which $\mathbfup{\gamma^\prime}=\infty$. In the top row $C=0\text{ (liquid inclusions)}, 0.3,1$ and in the bottom row $C= 3$, $10$, $100$. Note that the vertical axes have different scales on each panel to visualize the relevant range of behavior.  The percentage deviation $\Delta E_{rel}(\%)$ is a monotonically increasing function of $\mathbfup{\gamma^\prime}$ for inclusions as stiff as, or stiffer than, their host. However, when $C<1$, $\Delta E_{rel}(\%)$ decreases monotonically as $\mathbfup{\gamma^\prime}$ increases from $0$ to $\mathbfup{\gamma^\prime}_{cl}(C)$, vanishes at $\mathbfup{\gamma^\prime}=\mathbfup{\gamma^\prime}_{cl}(C)$, and increases monotonically above $\mathbfup{\gamma^\prime}_{cl}(C)$.}  
\label{deviationseveralC} 
\end{figure*}

Fig.\,\ref{figseveralC} shows the behavior of $E_{rel}$ as a function of $\phi$, for different values of $C\equiv \mu_1/\mu_2$. Firstly, we see softening behavior for $\mathbfup{\gamma^\prime} <\mathbfup{\gamma^\prime}_{cl}(C)$ (see Eq.\,(\ref{cloakingpoint})), and stiffening behavior for $\mathbfup{\gamma^\prime} >\mathbfup{\gamma^\prime}_{cl}(C)$. 
Secondly, in the  $C \rightarrow 0$ limit of the present theory, the dilute \cite{Stylesoft15,Style15} and non-dilute \cite{MSW3phase,MSWa} theories for liquid inclusions in soft solids are recovered quantitatively. 
Thirdly, we determine an exact condition for ``mechanical cloaking'' of the inclusions.  In particular, when the inclusions are softer than the surrounding host solid ($C<1$) the cloaking condition shows that 
$E_{rel} \equiv 1$ at $\mathbfup{\gamma^\prime}_{cl}(C)$ for all $\phi$. Moreover, because the cloaking radius $R_{cl}=\mathbfup{\gamma^\prime}^{-1}_{cl}L=\mathbfup{\gamma^\prime}^{-1}_{cl}(\gamma/E_2)$ is $\phi$-independent, this cloaking condition is a surprising generalization of that for liquid inclusions;  $C=0$ \cite{Stylesoft15,MSW3phase,MSWa}.  Fourthly, as may be intuitive, surface tension effects are not able to cloak the far-field signatures of inclusions that are stiffer than the matrix ($C\geq 1$).    Finally, Fig.\,\ref{figseveralC} provides some qualitative sense of how the $E_{rel}$ predictions deviate from the dilute theory obtained by linearizing in $\phi$ about $\phi=0$. Importantly, the dilute approximation defined in this manner differs from the dilute theory for liquid inclusions \cite{Style15,Stylesoft15,MSW3phase,MSWa}, which do not correspond to a linearization of $E_{rel}$, although these dilute models coincide in the small-$\phi$ limit. The percentage deviation $\Delta E_{rel}(\%)$ is shown in Fig.\,\ref{deviationseveralC}, but we note that this should not be compared with Fig.\,4 of \cite{MSW3phase}, where an appropriate, but different, dilute theory was used for comparison. Fig.\,\ref{deviationseveralC} demonstrates that $\Delta E_{rel}(\%)$ is a positive quantity, and that for modest values of $C$ (such as $C=0,0.3,1$) the dilute approximation  consistently breaks down beyond volume fractions of $\phi \approx 0.15$. In stark contrast, for larger values of $C$  (such as $C=3,10,100$), an accuracy of a few percent is only feasible in the dilute approximation up to volume fractions of $\phi \approx 0.1$.  The value of the calculations shown in Figs.\,\ref{figseveralC} and \ref{deviationseveralC} is the potential they provide for comparison with experiment and simulations; the dilute limit of the non-dilute theory for small to moderate values of the stiffness contrast $C$ offers a particularly robust regime.  

The $\phi$-independence exhibited by the cloaking condition (\ref{cloakingpoint}) has the following simple interpretation, valid for either liquid \cite{MSW3phase} or solid elastic inclusions. When shear modulus cloaking occurs, the materials in regions $i=2$ (matrix) and $i=3$ (composite effective medium) have identical shear modulus. Therefore, the continuity conditions on stress tensor and displacement vector at their common boundary $S^{ext}$ imply continuity of the shear strain as well. Hence, regions $i=2$ and $i=3$ behave as a single material with uniform shear properties, and thus the radial coordinate $R/\phi^{1/3}$ of their common boundary $S^{ext}$ (through which the volume fraction is expressed in the model) cannot affect the shear modulus of the composite.  For this reason, the cloaking condition is independent of $\phi$. Therefore, such an invariance of the cloaking condition is geometrically intrinsic to the 3-phase GSC approach, and provides an intriguing experimental target for actual composites with spatially-random distributions of many inclusions.  One might expect violations of the $\phi$-invariance of $\mathbfup{\gamma^\prime}_{cl}$ (or the cloaking radius $R_{cl}$), although for randomly distributed inclusions further study is needed to describe even simple qualitative details of the $\phi$-dependence.

\subsection{Macroscopic Equivalence and Surface Tension} \label{submacroequiv}

Next we show the (macroscopic) equivalence between an inclusion
embedded in an elastic solid with an isotropic interfacial
tension $\gamma$, and an ``equivalent" elastic inclusion with no
interfacial tension. Note that such a macroscopic mechanical equivalence differs from local micromechanical equivalence (see also \S \ref{simultaneousdetermination}). Thus, here we will calculate the \textit{equivalent} shear modulus $\mu^{\ast}$ of the actual inclusion.  We therefore compare Eshelby's results for the elastic moduli of a dilute composite with spherical elastic inclusions \footnote{The last two equations on pg.\,390, of Ref. \cite{Eshelby57}}; 
\begin{subequations} 
\begin{align} 
&\overline{K}^{dil}=\frac{K_2}{1-A^{-1} \phi},\quad A=\frac{1+\nu_2}{3(1-\nu_2)}\,, \label{bulkdiluteapprox}\\
\overline{\mu}^{dil}=\frac{\mu_2}{1+B\phi}&, \quad B=\frac{\mu^{\ast}-\mu_2}{(\mu_2-\mu^{\ast})\beta-\mu_2},\quad \beta=\frac{2}{15}\frac{4-5\nu_2}{1-\nu_2}\,,
\label{sheardiluteapprox} 
\end{align}
\end{subequations} 
to our expression (\ref{quadraticgeneral}) for the shear modulus of a composite containing incompressible inclusions in the dilute limit, which leads to
\be
\frac{\overline{\mu}^{dil}}{\mu_2}\approx 1+
\frac{5 [8 (-2 + 3 \mathbfup{\gamma^\prime}) + C (-3 + 19 C + 60 \mathbfup{\gamma^\prime})] }{48 + 
 89 C + 38 C^2 + 120 (1 + C) \mathbfup{\gamma^\prime}}\phi\;.
\label{shear_eqn}
\ee
Therefore, in Eqs. (\ref{sheardiluteapprox}) and (\ref{shear_eqn}) we equate different expressions for $\overline{\mu}^{dil}$ and thereby define a relative equivalent stiffness as $C^{\ast}\equiv \mu^{\ast}/\mu_2=(E^{\ast}/E_2)\vert_{\nu_2=1/2}$, and thus obtain
\be
C^{\ast}(\mathbfup{\gamma^\prime},C)=\frac{80 C + 95 C^2 + 192 \mathbfup{\gamma^\prime} + 300 C \mathbfup{\gamma^\prime}}{
 80 + 95 C + 72 \mathbfup{\gamma^\prime}}.
\label{equivalentshear}
\ee
Importantly we note that, although a small but {\em finite} value of the volume fraction $\phi\ll 1$ was assumed in the derivation, $C^{\ast}$ in Eq. (\ref{equivalentshear}) is {\em independent of} $\phi$.  
Additionally, having just evaluated $\mu^{\ast}$, we note that, by appealing to the equivalent inclusion concept, a Mori-Tanaka approach to the elastic characterization of our two-phase system can also be implemented. The equivalent moduli of an inclusion are easily evaluated analytically in the general isotropic case \footnote{Eq.\,(\ref{equivalentshear}) here, corresponding to the equivalent modulus in the simplest case of incompressible matrix.}, and in the dilute regime we can identify a single inclusion possessing surface tension with a single surface tension-free equivalent inclusion. Inside the latter inclusion, due to Eshelby's theorem \cite{Eshelby57}, the stress tensor would be uniform, and therefore the Mori-Tanaka estimate of the composite moduli in the non-dilute regime could be computed in direct analogy with what was done for liquid inclusions in Ref.\,\cite{MSWa}.
\subsection{Inequalities for Equivalent Stiffness and Relative Modulus}
Returning to the dilute approximation, our approach recovers the expression for $E^\ast$ of Style et al., \cite[][Eq.\,(9)]{Style15} for liquid inclusions ($C=0$), which is 
\be
\left(\frac{E^{\ast}}{E}\right)\vert_{C=0}=\frac{24{\gamma^\prime}}{10+9{\gamma^\prime}}=\frac{24\frac{L}{R}}{10+9\frac{L}{R}}.
\label{equivalentshearliquid}
\ee
In the small ($\mathbfup{\gamma^\prime} \rightarrow \infty$) and large ($\mathbfup{\gamma^\prime} \rightarrow 0$) inclusion limits, the equivalent stiffness asymptotes to the following two expressions;
\be 
C^{\ast}\stackrel  {\mathbfup{\gamma^\prime} \rightarrow \infty} \longrightarrow \;\frac{8}{3}+\frac{25}{6} C\;, \quad\textrm{and}\quad C^{\ast} \stackrel  {\mathbfup{\gamma^\prime} \rightarrow 0} \longrightarrow \; C+\frac{12}{5}\mathbfup{\gamma^\prime}\;,
\label{asymptotesequivalent}
\ee	
both in agreement with the predictions of Style et al. \cite{Style15} for the special case of liquid ($C=0$) inclusions. 

We emphasize that the surface tension and bulk elasticity contributions to the equivalent shear modulus $\mu^{\ast}$ of a single inclusion fulfill a kind of ``superadditivity".  Namely, Eq.(\ref{equivalentshear}) and its liquid case limit ($C=0$), Eq. (\ref{equivalentshearliquid}), imply that 
\be 
C^{\ast}>C+C^{\ast}\vert_{liquid}\;,
\label{superadditivity}
\ee 
for arbitrary values of $\mathbfup{\gamma^\prime}=L/R$ and $C=\mu_1/\mu_2$. Therefore, in the fully-incompressible case $\nu_1=\nu_2=1/2$, the interplay between surface tension and the bulk elasticity of a single elastic inclusion is such that the sum of their two separate contributions to the inclusion's equivalent stiffness underestimates the latter.   In the small $\mathbfup{\gamma^\prime}$ limit we can rewrite Eq. (\ref{superadditivity}) in terms of the Young's moduli as
\be 
E^{\ast}>E_1 + \frac{12}{5} \frac{{\gamma}_0 (1-\epsilon)}{R}\;,
\label{superadditivityE}
\ee 
where $\epsilon$ is a small parameter characterizing the a surfactant dependent reduction in the ``bare'' liquid droplet/matrix interfacial tension ${\gamma}_0$, which we envision controlling.  This is a mean field description of the equivalent modulus of the inclusion ($E^{\ast}$)  in terms of the bulk value ($E_1$) of the elastic inclusion plus the ``Laplace--pressure'' associated with the liquid droplet.  Firstly, in the limit of vanishing surface tension ($\epsilon\rightarrow1^{-}$) the inequality is violated and ``bulk-coexistence'' prevails; $E^{\ast} = E_1$.  One can think of this in the same sense as the coexistence of two bulk thermodynamic phases reached when a droplet of one phase either becomes arbitrarily large or has no surface tension to maintain the geometrical configuration.  The analogy is strictly heuristic.  As $\epsilon\rightarrow0^{+}$ the equivalent elastic inclusion of the liquid droplet is always less stiff than its actual elastic counterpart; the pressure in a droplet is less than the normal stress in an equivalent elastic inclusion via the generallized Young-Laplace equation.  Hence the inequality.

Finally, we define a \textit{residual} equivalent stiffness $C^{\ast}_{RES}\equiv [C^{\ast}-C-C^{\ast}\vert_{liquid}]$, and from Eq. (\ref{superadditivity}) it must hold that:
\be
C^{\ast}_{RES}=\frac{2052 C {\mathbfup{\gamma^\prime}}^2}{(10 + 9 \mathbfup{\gamma^\prime})(80 + 95 C + 72 \mathbfup{\gamma^\prime})}>0
\ee
The magnitude of the inequality (\ref{superadditivity}) increases  with decreasing inclusion size, and thus the \textit{residual} equivalent stiffness peaks at a maximum; $\textrm{max}\{C^{\ast}_{RES}\}=19C/6$ for $\mathbfup{\gamma^\prime}\rightarrow \infty$ and vanishes as $\sim[1/R^2]$ for $\mathbfup{\gamma^\prime}\rightarrow 0$, as suggested by Eqs.\,(\ref{asymptotesequivalent}). Although surface and bulk effects have been shown to ``decouple" in the limit of large radius $R$, they maintain a non-trivial coupling in the ideal limit of rigid spheres, $C \rightarrow \infty$, leading to
\be
C^{\ast}_{RES}\stackrel  {C \rightarrow \infty} \longrightarrow \;\frac{108 {\mathbfup{\gamma^\prime}}^2}{5(10 + 9 \mathbfup{\gamma^\prime})}\;.
\ee

Conversely, the surface tension and bulk elasticity contributions to the relative stiffness deviation of the composite in the dilute regime, $\Delta E_{rel} \equiv (E_3-E_2)/E_2$, fulfill a kind of ``subadditivity", viz., 
\be 
\Delta E_{rel}<\Delta E_{rel}\vert_{liquid}+\Delta E_{rel}\vert_{bulk},
\label{subadditivity}
\ee 
for any value of $\mathbfup{\gamma^\prime}=L/R$ and $C=\mu_1/\mu_2$. In Eq. (\ref{subadditivity}), all terms refer to predictions in the dilute regime and thus $\Delta E_{rel}\vert_{liquid}$ is 
\be
\Delta E_{rel}\vert_{liquid}\equiv E_{rel}(C=0,\mathbfup{\gamma^\prime})-1=\frac{5}{3}\frac{3\mathbfup{\gamma^\prime}-2}{2+5\mathbfup{\gamma^\prime}}\phi\;,
\ee
and $\Delta E_{rel}\vert_{bulk}$ is derived from Eqs.(\ref{bulkdiluteapprox}) and (\ref{sheardiluteapprox}) and is
\be
\Delta E_{rel}\vert_{bulk}\equiv E_{rel}(C,\mathbfup{\gamma^\prime} =0)-1=\frac{C-1}{1+\frac{2}{5}(C-1)}\phi\;.
\ee
Now, defining a \textit{residual} relative modulus deviation $\Delta E^{RES}_{rel}\equiv [\Delta E_{rel}-\Delta E_{rel}\vert_{bulk}-\Delta E_{rel}\vert_{liquid}]$, it holds to first order in $\phi$ that
\begin{multline}
\Delta E^{RES}_{rel}=-\frac{20 C \mathbfup{\gamma^\prime}\phi}{3 (3 + 2 C) (2 + 5 \mathbfup{\gamma^\prime})} \;\times \\
 \frac{ 384 + 345 \mathbfup{\gamma^\prime} +
    8 C (73 + 19 C + 60 \mathbfup{\gamma^\prime})}{
  48 + 89 C + 38 C^2 + 
    120 (1 + C) \mathbfup{\gamma^\prime}}\;<0\;.
\label{negativeresidual}
\end{multline}
A physical interpretation of the inequality expressed in Eqs.\,(\ref{subadditivity}) and (\ref{negativeresidual}) rests on the relative influence of interfacial versus elastic forces. Whence, the inequality can be said to reside in the fact that the contribution of surface tension to the stiffening of the composite is less than it would be -- all else being equal -- in the liquid inclusion case. This is in agreement with the general understanding that the impact of surface tension on the macroscopic properties of a composite is smaller for overall stiffer composites; a composite with elastic inclusions instead of liquid inclusions -- all else being equal -- is definitely a stiffer composite. 
In the large inclusion-limit $\mathbfup{\gamma^\prime}\rightarrow 0$, the residual deviation $\Delta E^{RES}_{rel}$ vanishes as $\sim[1/R]$.   The competitive interaction between bulk and surface-related contributions to the modulus deviation, represented by (\ref{subadditivity}), peaks at the minimum value $\textrm{min}\{\Delta E^{RES}_{rel}\}=-32C\left(16+19 C\right)/\left[3(3+2C)(41+38C)\right]$ for $\mathbfup{\gamma^\prime}=(64+76C)/45$, and asymptotes at $\Delta E^{RES}_{rel}=-C\left(23+32C\right)/\left[(6(1+C)(3+2C)\right]$ in the small inclusion limit $\mathbfup{\gamma^\prime} \rightarrow \infty $. The inequalities (\ref{superadditivity}) and (\ref{subadditivity}) have complementary but opposite character, and refer to inclusion equivalent stiffness and composite effective stiffness respectively.  Together they demonstrate that interfacial tension and bulk inclusion elasticity both contribute to the composite elastic properties through complex mutual interactions.

\subsection{Measuring surface tension and inclusion stiffness}\label{simultaneousdetermination}

Here we discuss a general experimental setting for the simultaneous determination of interfacial tension and bulk elasticity, based on measurements of global and local mechanical effects. For example, combined measurements of a local property, such as the effective inclusion strain $\varepsilon_{inc}$, and a bulk global property, such as the composite stiffness $E_{3}$, are sufficient to infer both the surface tension parameter $\mathbfup{\gamma^\prime}$ and the inclusion/matrix stiffness contrast $C$.  Thus, in the dilute regime Eqs.\,(\ref{diluteinclusionstrain}) and (\ref{shear_eqn}) express the dependence of two measurable dimensionless quantities, the relative inclusion strain $\varepsilon_{inc}^{rel}\equiv (\varepsilon_{inc}/2)/\varepsilon_{zz}^0=\varepsilon_{inc}/(4\varepsilon_{A}^0)$ and the relative reinforcement slope $S \equiv [E_3(\phi)-E_3(\phi=0)]/(E_2\phi)$, on $\mathbfup{\gamma^\prime}$ and $C$. These relationships can be combined to express $C$ and $\mathbfup{\gamma^\prime}$ as functions of the experimental parameters $S$ and
$\varepsilon_{inc}^{rel}$ defined above as follows; 
\be
C = -\frac{2 \left(8 \varepsilon^{rel}_{inc}+5 S-5\right)}{19 \varepsilon^{rel}_{inc}+10
   S-25}\; \textrm{~and}
\ee
\be
\mathbfup{\gamma^\prime} = \frac{5 (S+7) \left(5 \varepsilon^{rel}_{inc}+2 S-5\right)}{12 \varepsilon^{rel}_{inc} \left(19 \varepsilon^{rel}_{\text{inc}}+10 S-25\right)}\;.
\ee
\subsection{Comments on the auxetic case}
For an incompressible matrix, spherical liquid inclusions with radii less than the cloaking radius, $R < R_{cl} =(3/2) L$, stiffen the matrix \cite{Style15, Stylesoft15}.
However, for a compressible matrix having Poisson's ratio $0<\nu_2<1/2$, the cloaking radius $R_{cl}=9L$ (at $\nu_2=0$) is larger, see Eq.\,(19) of \cite{Stylesoft15}.
Similarly, for auxetic  matrix materials ($\nu_2<0$), the cloaking radius increases further, diverging as $\nu_2 \rightarrow -1/13$. Accordingly, auxetic matrices in the physical range $-1<\nu_2<-1/13$ are stiffened by liquid inclusions of {\em any size}, independent of the elastocapillary length scale. 

We note that in the regime where $\nu_2$ approaches $-1/13$, even at length scales much larger than $L$, interfacial effects dominate over bulk elasticity in terms of deviations of moduli, although such deviations appear small when compared with matrix moduli.
Indeed, at $\nu_2=-1/13 $, Eshelby's theory would predict cloaking, however the stiffening of the matrix can be entirely ascribed to surface tension (for any value of $R$), which is not included in Eshelby's theory. 
This argument can be repeated for solid inclusions, with the effect of  raising in the threshold value of $\nu_2$ to be greater than $-1/13$.

\section{Conclusions} \label{Conclusions}

A complete mechanical characterization of inclusion-matrix composites is of great interest to the soft condensed matter and life science communities. Previously we examined the effects of surface tension in the bulk behavior of soft composite solids hosting liquid inclusions \cite{Style15, Stylesoft15, MSW3phase, MSWa}, to find that for inclusions smaller (larger) than the elastocapillarity length the matrix stiffened (softened), thereby finding novel behavior relative to the classical theory of Eshelby, which predicts only softening. Here we have extended and generalized a three-phase generalized self-consistent approach \cite{MSW3phase} to consider isotropic, linear-elastic spherical inclusions, whose surface tension with the host solid is accounted for.  In the case of an incompressible matrix, we calculate the composite shear modulus (stiffness) as a function of the dimensionless parameters $\phi$ (inclusion concentration), $C=\mu_1/\mu_2$ (stiffness contrast between the inclusions and the matrix) and $\mathbfup{\gamma^\prime}=\gamma/(R E_2)$ (an elastocapillary number), as well as the effective strain and the deformed shape of the inclusions themselves.  We verified that our approach yields the known results in the limiting cases in which the inclusions are liquid ($C=0$) and rigid ($C=\infty$).

The span of the parameters $C$ and $\mathbfup{\gamma^\prime}$ in Fig.\,\ref{figseveralC} were chosen to display the range  of regimes that can occur in biological or industrially manufactured soft composites.
Remarkably, even for inclusions one hundred times stiffer than their host matrix, the effect of surface tension on the effective Young's modulus is substantial; particularly so at high concentrations $\phi$. This is compatible with realistic values of the parameters found in soft composites. An estimate of the surface tension of a wide range of soft solids is of the order of $\gamma \sim$ 50mN/\,m (e.g., see Refs.\, \cite{deGennes10} and \cite{Dalal87}); considering matrix materials having Young's moduli in the range of $E_2 \sim 10 $kPa (compatible with several kinds of human and animal tissues \cite{Buxboim10,Tyler12}) with nanoinclusions of radius $R=50$ nm and a hundred times stiffer than their host, the regime described by the  $\mathbfup{\gamma^\prime}=100$ curve in the bottom right panel ($C=$100) of Fig.\,\ref{figseveralC} is appropriate. Thus, according to our model, in this case surface tension is responsible for a significant fraction of the composite stiffness, even for such a large stiffness contrast.  Indeed, at inclusion volume fractions of  $\phi=0.45$ and $\phi=0.50$, the respective normalized Young's moduli $E_{rel}$ of the composites are $\sim 9 \%$ and $\sim 15 \%$ higher than the interface-free counterpart. Surface tension's influence on the Young's modulus {\em increases} when inclusion radius, stiffness contrast, matrix stiffness, or volume fraction ($1-\phi$) of the matrix component {\em decrease}. As is the case for liquid inclusions, we find a mechanical cloaking radius, which by definition leads to an effective response of the composite that is independent of $\phi$, and hence surface tension effectively cloaks the far-field influence of inclusions. In contrast to the liquid case, rather than the cloaking value of $\mathbfup{\gamma^\prime}$ being a constant, $\mathbfup{\gamma^\prime}_{cl} = 2/3$, here it is a function of the stiffness contrast $C$, as shown in the dilute limit by Eq. (\ref{cloakingpoint}). Moreover, when this functional condition is met, the inclusions will stretch less than the host material.   We compared and contrasted these cases briefly in the case of auxetic matrix materials.

We introduced the two notions of relative equivalent stiffness of the inclusions and relative stiffness deviation of the composite, and demonstrated that they possess properties that we call ``superadditivity'' and ``subadditivity'', respectively. These two inequalities demonstrate a non-trivial interaction between the interfacial tension and bulk elasticity of the inclusions, which determine the composite effective stiffness.  This suggests the possible experimental or numerical determination of the surface tension and the stiffness contrast parameters through simultaneous measurement of the relative inclusion strain $\varepsilon_{inc}^{rel}$ and the relative reinforcement slope $S$.  Clearly, such measurements have a broad range of applications. This study on the shear modulus of two-phase soft composites, suggests a number of natural extensions, such as finite deformations, recently addressed for liquid spherical inclusions using finite-element modeling \cite{Wang16}, as well as the behavior of composites with spheroidal- or arbitrarily-shaped  inclusions. 

The so-called deformation amplification approach, originally proposed for rigid inclusions \cite{Bergstrom99}, was recently extended to the case of liquid spherical inclusions by Wang and Hennan \cite{Wang16} using numerical simulations. They collapsed on a single curve the nominal strain response of liquid inclusion-matrix composites versus the nominal applied stress (normalized by the infinitesimal shear modulus of the composite), for different values of $\phi$ and $\mathbfup{\gamma^\prime}$ 
 \cite{Wang16}. This suggests an intriguing possibility of characterizing the large deformation behavior of composites with elastic inclusions.  Indeed, this collapse of a family of constitutive relations to a single curve would imply that the homogenized composite material exhibits a hyperelastic behavior, identical in functional form to that of the matrix material itself.

Finally, returning to the framework of infinitesimal deformations, we speculate that randomly oriented inclusions of complex shapes (e.g., spheroidal, rod-like, or arbitrarily-shaped) would have a larger stiffening/softening effect than if they were spherical inclusions of identical volume, surface tension, volume fraction and stiffness. Considering that such is the case for composites in which interfacial effects are negligible, we envisage it to be approximately fulfilled even far from the bulk elasticity-dominated regime. It is hoped that our calculations may provide a baseline for future experimental and numerical studies.  

\section{Acknowledgments}\label{Acknowledgments}
The authors thank R.W. Style for his encouragement to examine this system.  FM and JSW acknowledge Swedish Research Council Grant No. 638-2013-9243.  JSW also acknowledges a Royal Society Wolfson Research Merit Award.  
\appendix
\section{Coefficients in Equation (\ref{quadraticgeneral})  \label{AppendixA}}
The coefficients in equation (\ref{quadraticgeneral}) for the case of an incompressible matrix ($\nu_2=1/2$) are as follows
\begin{widetext}
\be
\left\{
\begin{array}{l}
a_0(\alpha,C)=\alpha ^{10} \left(-722 C^2-1691 C-912\right)+\alpha ^7 \left(-4275 C^2+675 C+3600\right)+\alpha ^5 \left(6384 C^2-1008 C-5376\right)\\
\phantom{.................}+\alpha ^3 \left(-3800 C^2+200 C+3600\right)-912 C^2+1824 C-912\\\\
a_1(\alpha,C)=\alpha ^{10} \left(114 C^2+267 C+144\right)+\alpha ^7 \left(4750 C^2-750 C-4000\right)+\alpha ^5 \left(-12768 C^2+2016 C+10752\right)\\ \phantom{.................}+\alpha ^3 \left(7600 C^2-400
   C-7200\right)+304 C^2-608 C+304\\\\
a_2(\alpha,C)=\alpha ^{10} \left(608 C^2+1424 C+768\right)+\alpha ^7 \left(-3800 C^2+600 C+3200\right)+\alpha ^5 \left(6384 C^2-1008 C-5376\right)\\\phantom{.................}+\alpha^3 \left(-3800 C^2+200 C+3600\right)+608 C^2-1216 C+608
\end{array}
\right. , 
\ee
and 
\be
\left\{
\begin{array}{l}
b_0(\alpha,C)=-1520 \alpha ^{10} (C+1)-1800 \alpha ^7 (5 C+2)+13440 \alpha ^5 C+1600 \
\alpha ^3 (2-5 C)-1920 (C-1)\\\\
b_1(\alpha,C)=240 \alpha ^{10} (C+1)+2000 \alpha ^7 (5 C+2)-26880 \alpha ^5 C+3200 \
\alpha ^3 (5 C-2)+640 (C-1)\\\\
b_2(\alpha,C)=1280 \alpha ^{10} (C+1)-1600 \alpha ^7 (5 C+2)+13440 \alpha ^5 C+1600 \
\alpha ^3 (2-5 C)+1280 (C-1)
\end{array}
\right.. 
\ee

\end{widetext} 

\section{Shape of the inclusions under uniaxial stress \label{AppendixB}}

We determine the shape of each incompressible inclusion embedded in an incompressible matrix under uniaxial stress.  Here, Eqs. (\ref{radialansatz}) and (\ref{polaransatz}) reduce to
\be 
\frac{u_r(1,\theta)}{R}=(6\mathcal{A}_2+2\mathcal{B}_2+6\mathcal{C}_2-3\mathcal{D}_2)\;\mathcal{P}_2(\cos \theta)\;,
\label{dropshaper}
\ee
and
\be
\frac{u_\theta(1,\theta)}{R}=(5\mathcal{A}_2+\mathcal{B}_2+\mathcal{D}_2)\;\frac{d \mathcal{P}_2(\cos \theta)}{d \theta}\;, 
\label{dropshapetheta}
\ee
respectively.  Finally, we note that the coefficients $\{f_i\}$ in Eqs. (\ref{radshape}) and (\ref{thetashape}) are as follows:
\begin{widetext}
\be
\left\{
\begin{array}{l}
f_1= \left[(19 C+16)\left(16  \mu _{\text{rel}}+19\right)\alpha ^7+336 (C-1)\left( \mu _{\text{rel}}-1\right)\alpha ^2 -640 (C-1) (\mu _{\text{rel}}-1)\right]/2 \\\\
f_2= \left(38 C^2+89 C+48\right) \left(48 \mu _{\text{rel}}^2 +89 \mu _{\text{rel}}+38\right)\alpha ^{10}-150 \left(19 C^2-3 C-16\right) \left(4 \mu_{\text{rel}}^2-\mu _{\text{rel}}-3\right)\alpha ^7 \\
\phantom{.......}+336 \left(19 C^2-3 C-16\right) \left(3 \mu _{\text{rel}}^2-\mu _{\text{rel}}-2\right)\alpha ^5-200 \left(19 C^2-C-18\right) \left(3 \mu _{\text{rel}}^2-\mu _{\text{rel}}-2\right)\alpha ^3+1824 (C-1)^2 (\mu _{\text{rel}}^2-2\mu _{\text{rel}}+1)\\\\
f_3=120 \left[ (C+1)\left(48 \mu _{\text{rel}}^2+89 \mu _{\text{rel}}+38\right)\alpha ^{10}-15(5C+2) \left(4 \mu_{\text{rel}}^2-\mu _{\text{rel}}-3\right)\alpha^7 \right.\\
\phantom{.......}+ 168 C\left( 3 \mu_{\text{rel}}^2-1 \mu_{\text{rel}}-2 \right) \alpha^5 - 20(5 C-2)\left(3 \mu_{\text{rel}}^2- \mu _{\text{rel}}-2 \right) \alpha ^3 \left. + 48(C-1)    (\mu_{\text{rel}}^2-2 \mu _{\text{rel}}+1) \right]\\\\
f_4=(19 C+16) \left(16 \mu _{\text{rel}}+19\right)\alpha^7 - 224 (C-1) \left(\mu_{\text{rel}}-1\right)\alpha^2 -80 (C-1) \left(\mu _{\text{rel}}-1\right)\\\\
f_5=24(16 \mu _{\text{rel}}+19) \alpha^7 -1344 \left(\mu _{\text{rel}}-1\right)\alpha^2 + 960 \left(\mu _{\text{rel}}-1\right)
\end{array}
\right. . 
\ee
\end{widetext}

\eject

\end{document}